\documentclass[envcountsame,reqno]{llncs}  

\usepackage{amssymb}
\usepackage{mathtools}   
\usepackage{enumerate}
\usepackage{xspace}
\usepackage{verbatim} 

\usepackage{braket}  

\usepackage{tikz}
\usetikzlibrary{calc,positioning}

\usepackage{letltxmacro}  
\LetLtxMacro{\Standardexists}{\exists}    
\renewcommand*{\exists}[1]{\Standardexists #1 \,}
\LetLtxMacro{\Standardforall}{\forall}
\renewcommand*{\forall}[1]{\Standardforall #1 \,}

\newcommand{\Bmc}{\mathcal{B}}

\newcommand{\Kmc}{\mathcal{K}}
\newcommand{\Lmc}{\mathcal{L}}

\newcommand{\Pmc}{\mathcal{P}}

\newcommand{\N}{\mathbb{N}}
\newcommand{\Z}{\mathbb{Z}}
\newcommand{\Q}{\mathbb{Q}}

\newcommand{\Struc}[1]{\mathcal{#1}}
\newcommand{\CG}[1]{\ensuremath{\mathbb{#1}}} 
\newcommand{\Func}{\mathcal{F}}
\newcommand{\VarEl}{\mathsf{Var}_1}
\newcommand{\VarSet}{\mathsf{Var}_2}

\newcommand{\image}{\mathop{\mathsf{im}}}
\newcommand{\dom}{\mathop{\mathsf{dom}}}

\newcommand{\Bool}{\mathsf{Bool}}
\newcommand{\Bound}{\mathsf{B}}
\newcommand{\CTL}{\mathsf{CTL}}
\newcommand{\LTL}{\mathsf{LTL}}
\newcommand{\ECTL}{\mathsf{ECTL}^*}

\newcommand{\Ex}{\mathsf{E}}
\newcommand{\homom}{\preceq}
\newcommand{\Prop}{\mathsf{P}}
\newcommand{\SAT}{\mathsf{SAT\text{-}ECTL}^*}

\newcommand{\MSO}{\ensuremath{\mathsf{MSO}}}
\newcommand{\WMSO}{\ensuremath{\mathsf{WMSO}}}
\newcommand{\WMSOB}{\ensuremath{\mathsf{WMSO\!+\!B}}}

\newcommand{\reach}{\mathsf{reach}}
\newcommand{\suc}{\mathsf{S}}
\newcommand{\arity}[1]{\mathsf{ar}({#1})}
\newcommand{\EHD}{\textsf{EHD}\xspace}
\newcommand{\road}{\mathsf{road}}

\newcommand*{\Fraisse}[0]{Fra\"iss\'e\xspace}

\renewcommand{\phi}{\varphi}

\newcommand{\Tree}{\Struc{T}_{\infty}}

\newcommand{\prefixCl}{{\downarrow}}

\newcommand{\inA}[1]{\varphi_{{#1}}}

\newcommand{\inc}{\mathrel{\bot}}

\newcommand{\TITEL}{Satisfiability of $\ECTL$ with tree constraints}
\newcommand{\AUTHOR}{Claudia Carapelle\inst{1},  Shiguang Feng\inst{1}, Alexander
  Kartzow\inst{2} and Markus Lohrey\inst{2}}
\newcommand{\KEYWORDS}{satisfiability, ECLT* with constraints, trees,
  semi-linear orders}

\usepackage{hyperref}

\hypersetup{
colorlinks=false, 
pdfauthor={\AUTHOR},
pdftitle={\TITEL},
pdfkeywords={\KEYWORDS}
}

\begin{document}

\title{\TITEL\thanks{This work is supported by the DFG Research Training
    Group 1763 (QuantLA) and the DFG
    research project GELO.}}

\author{\AUTHOR}

\institute{Institut f\"ur Informatik, Universit\"at Leipzig, Germany\and
Department f\"ur Elektrotechnik und Informatik,
  Universit\"at Siegen, Germany}

\maketitle

\begin{abstract}
  Recently, we have shown that satisfiability for $\ECTL$ with 
  constraints over $\mathbb{Z}$ is decidable using a new technique. 
  This approach reduces the 
  satisfiability problem of $\ECTL$ with constraints over some
  structure $\Struc{A}$ (or class of structures) to the problem whether $\Struc{A}$ has a
  certain model theoretic property that we called \EHD (for ``existence of homomorphisms is decidable''). Here we apply this approach to
  concrete domains that are tree-like and obtain several results. 
  We show that satisfiability of $\ECTL$ with constraints is decidable over 
  (i) semi-linear orders (i.e., tree-like structures where
    branches form arbitrary linear orders),
  (ii) ordinal trees (semi-linear orders where the branches form
    ordinals),
  and (iii) infinitely branching trees of height $h$ for each fixed $h\in\N$.
  We prove that all these classes of structures have the property \EHD. 
  In contrast, we introduce Ehrenfeucht-\Fraisse-games for $\WMSOB$ 
  (weak $\MSO$ with the bounding quantifier) and use them to show 
  that the  infinite (order) tree does not have property \EHD.
  As a consequence, a different approach has to be taken in order to
  settle the question whether satisfiability of $\ECTL$ (or even
  $\LTL$) with constraints over the infinite (order) tree  is decidable. 
\end{abstract}

\section{Introduction}

Temporal logics like $\mathsf{LTL}$, $\CTL$ or $\CTL^*$ are nowadays
standard languages for specifying system properties in
verification. These logics are interpreted over node labeled graphs (Kripke structures), where
the node labels (also called atomic propositions) represent abstract
properties of a system. Clearly, such an abstracted system state does
in  general not contain all the information of the original system state.
Consider for instance a program that manipulates two integer variables
$x$ and $y$.
A useful abstraction might be to introduce atomic
propositions $v_{-2^{32}}, \ldots, v_{2^{32}}$ for $v \in \{x,y\}$,
where the meaning of $v_k$ for $-2^{32} < k < 2^{32}$ is that the
variable $v\in \{x,y\}$ currently holds the value $k$,  and 
$v_{-2^{32}}$  (resp., $v_{2^{32}}$) means that 
the current value of $v$ is at most $-2^{32}$ (resp., at least $2^{32}$). It is evident that such
an abstraction might lead to incorrect results in model-checking.

To overcome this problem, extensions of temporal logics with constraints
have been studied. In this setting, a model of a formula is not only a
Kripke structure but a Kripke structure where every node is assigned
several values from some fixed structure
$\Struc{C}$ (called a concrete domain). The logic is then enriched in
such a way that it has access to the relations of the concrete
domain. For instance, if $\Struc{C} = ( \Z, =)$ then every node of the
Kripke structure gets assigned several integers and the logic
can compare  the values assigned to neighboring nodes for equality. 

In a recent paper \cite{CarapelleKL13} we introduced a new
method (called \EHD-method in the following) which
shows decidability of the satisfiability problem for $\CTL^*$
extended by local constraints over the integers. The latter logic was first
introduced in \cite{DemriG08}. In \cite{CKL14} we extended the \EHD-method
to \emph{extended 
computation tree logic} ($\ECTL$) with constraints over the integers (a
powerful temporal logic that properly extends $\CTL^*$) and  proved that  satisfiability is still decidable. 
This result greatly improves the
partial results on fragments of $\CTL^*$ obtained by Bozzelli, Gascon
and Pinchinat \cite{BozzelliG06,BozzelliP14,Gascon09}.

The idea of the \EHD-method is as follows. 
Let $\Struc{C}$ be any concrete domain over a relational signature $\sigma$. 
Satisfiability of $\ECTL$ with constraints over $\Struc{C}$ is
decidable if $\Struc{C}$ has the following two properties:
\begin{enumerate}
\item The structure $\Struc{C}$ is negation-closed, i.e., the complement of any
  relation $R \in \sigma$ is definable in positive existential
  first-order logic.
\item There is a logic $\Lmc$ with certain properties (listed below) and an
  $\Lmc$-sentence $\varphi$ characterizing homomorphism to
  $\Struc{C}$ in the sense that for any countable $\sigma$-structure
  $\Struc{A}$ there is a homomorphism from $\Struc{A}$ to $\Struc{C}$
  if and only if $\Struc{A}\models \varphi$.
\end{enumerate}
For the candidate logics $\Lmc$ we need the following properties:
\begin{enumerate}
\item Satisfiability of a given $\Lmc$-sentence
  over the class of infinite node-labeled trees is decidable.
\item  $\Lmc$ is closed under boolean combinations with monadic
  second-order formulas ($\MSO$).
\item $\Lmc$ is compatible with one dimensional first-order
  interpretations and with the $k$-copy operation, see \cite{CKL14}.
\end{enumerate}
The most powerful logic with these properties that we are aware of
is the set of all Boolean combinations of $\MSO$-formulas
and $\WMSOB$-formulas (briefly $\Bool(\MSO,\WMSOB)$), where 
$\WMSOB$ is  weak monadic second-order
logic with the bounding quantifier. The bounding quantifier allows to express
that there is a bound on the size of finite sets satisfying a certain property.
Satisfiability of $\WMSOB$ over infinite node-labeled trees was shown to be decidable
in \cite{BojanczykT12} (in contrast, decidability of full monadic second-order logic
with the bounding quantifier over infinite node-labeled trees cannot
be proved in ZFC \cite{BGMS14}).
In \cite{CarapelleKL13} we proved that the existence of a homomorphism into
$(\mathbb{Z},<,=)$ can be expressed in $\Bool(\MSO,\WMSOB)$ (or even in 
$\WMSOB$).

These results gave rise to the hope that the \EHD-method
could also be fruitfully applied to other concrete domains. An interesting candidate 
in this setting is the full infinitely branching infinite (order) tree
$\Tree=(\N^*, <, \inc, =)$,
where $<$ denotes the prefix order on $\N^*$ and
$\inc$ denotes the incomparability relation with respect to
$<$ (this structure is negation-closed, which is the reason for adding the  incomparability relation $\inc$). 
Unfortunately, this hope is destroyed by one of the main results
of this work: 
\begin{theorem}
  \label{thm:MainTheorem2}
  There is no $\Bool(\MSO, \WMSOB)$-sentence $\psi$ such that 
  for every countable structure $\Struc{A}$ (over the signature 
  $\{<,\inc,=\}$) we have: $\Struc{A} \models \psi$ 
  if and only if there is a homomorphism from $\Struc{A}$ to $\Tree$.
\end{theorem}
This result is shown using a suitable
Ehrenfeucht-\Fraisse-game.

Theorem~\ref{thm:MainTheorem2} shows that the \EHD-method cannot be applied
to the concrete domain $\Tree$. Of course, this
does not imply that  satisfiability for $\ECTL$ with constraints over $\Tree$
is undecidable, which remains an open problem (even for 
$\LTL$ instead of $\ECTL$).  In fact, we 
still conjecture that satisfiability for $\ECTL$ with constraints over $\Tree$
is decidable and we support this
conjecture by fruitfully applying the \EHD-method to other tree-like
structures, such as semi-linear orders, ordinal
trees, and infinitely branching trees of a fixed height. 
{\em Semi-linear orders} are partial orders that are
tree-like in the sense that for every element $x$
the set of all smaller elements  form a linear
suborder. If this linear suborder is an ordinal (for every $x$) then
one has an {\em ordinal tree}. 
Ordinal trees are widely studied in descriptive set theory and
recursion theory.  Note that a tree is a connected semi-linear order
where for every element the set of all smaller elements is a finite
linear order. 

In the integer-setting from \cite{CarapelleKL13,CKL14}, we investigated satisfiability for $\ECTL$-formulas with
constraints over one fixed structure (integers with additional relations). 
For semi-linear orders and ordinal trees it is more natural to consider satisfiability with respect to a class
of concrete domains $\Gamma$ (over a fixed signature $\sigma$):
The question becomes, whether for a given constraint $\ECTL$
formula $\phi$ there is a concrete domain $\Struc{C} \in \Gamma$ such that $\phi$ is satisfiable by some model
with concrete values from $\Struc{C}$?\footnote{If the class $\Gamma$ is closed under taking induced substructures
(which is the case for our classes)
then one can restrict $\Gamma$ to its countable members.}
If a class $\Gamma$ has a universal structure\footnote{A structure
 $\Struc{U}$ is universal for a class $\Gamma$ if there is a
 homomorphic embedding of every structure from $\Gamma$ into
 $\Struc{U}$ and $\Struc{U}$ belongs to $\Gamma$.}
$\Struc{U}$, then  satisfiability with respect to the class $\Gamma$ 
is equivalent to satisfiability with respect to $\Struc{U}$ because
one easily shows that a formula $\phi$ has a
model with some concrete domain  from $\Gamma$ if and only if it has a
model with concrete domain $\Struc{U}$. 
A typical class with a universal model is the class of all countable
linear orders, 
for which $(\mathbb{Q},<)$ is universal. Similarly, for the class of
all countable  
trees the infinitely branching infinite tree as well as the binary
infinite tree are universal. 
In the appendix we construct a universal countable semi-linear order.
Since this particular universal structure appears to be less natural than 
$(\mathbb{Q},<)$ or the infinite binary tree, we have decided to formulate
our decidability result for the class of all semi-linear orders.
Moreover, there is no universal structure for the class of
countable ordinal trees (for a similar reason as the one showing that
the class of countable ordinals does not contain a universal structure).

Application of  the \EHD-method to semi-linear orders and ordinal trees 
gives the following decidability
results.
\begin{theorem}
  \label{thm:MainTheorem1}
  Satisfiability of $\ECTL$-formulas with
  constraints over each of the following classes is decidable:
  \begin{enumerate}[(1)]
  \item the class of all  semi-linear orders,
  \item the class of all  ordinal trees, and
  \item for each $h\in \N$, the class of all  order trees
    of height $h$. 
  \end{enumerate}
\end{theorem}
Concerning computational complexity, let us remark that in
\cite{CarapelleKL13,CKL14} we did not present an upper bound on the
complexity of our decision procedure.  The reason for this is that the
authors of \cite{BojanczykT12} do not proof an upper bound for the
complexity of satisfiability of $\WMSOB$ over infinite trees, even in
the case that the input formula has bounded quantifier depth (and it
is not clear how to obtain such a bound from the proof of
\cite{BojanczykT12}. Here, the situation is slightly different. Our
applications of the $\EHD$-method for the proof
Theorem~\ref{thm:MainTheorem1} do not need the bounding quantifier,
and classical $\WMSO$ (for semi-linear orders) and $\MSO$ (for ordinal
trees and trees of bounded height) suffice. Moreover, the formulas
that express the existence of a homomorphism have only small
quantifier depth (at least for semi-linear orders and ordinal trees;
for trees of bounded height, the quantifier depth depends on the
height).  This fact can be exploited and yields a 
triply exponential upper bound on the time complexity in (1) and (2)
from Theorem~\ref{thm:MainTheorem1} (this bound does
not match the doubly exponential lower bound inherited from
the satisfiability problem of $\ECTL$ without constraints). 
We skipped the proof details, since we conjecture the exact complexity
to be doubly exponential.

The paper is organized as follows: Section~\ref{sec-prel} introduced
the necessary machinery concerning Kripke structures, tree-like
partial orders and the logics $\MSO$, $\WMSOB$ and $\ECTL$ with
constraints. Moreover, we 
explain the \EHD-method (more details can be found in \cite{CKL14}).
Using the \EHD-method, Theorem~\ref{thm:MainTheorem1} is proved in
Sections~\ref{sec-decidable-semilinear} (for semi-linear orders),
\ref{sec-ordinal-trees} (for ordinal trees) and
\ref{sec-bounded-height} (for trees of bounded height $h$).
Section~\ref{sec-trees} introduces an Ehrenfeucht-\Fraisse-game for
$\WMSOB$ and uses this game to prove Theorem~\ref{thm:MainTheorem2}.

\section{Preliminaries} \label{sec-prel}

In this section we introduce basic notations concerning Kripke
structures, various classes of tree-like 
structures, and the logics  $\MSO$, $\WMSOB$, and $\ECTL$ with
constraints.

\subsection{Structures}\label{sec-struc}

Let $\Prop$ be a countable set of (atomic) propositions.  A Kripke
structure (over $\Prop$) is a triple $\Kmc = (D,\to,\rho)$, where:
\begin{itemize}
\item $D$ is an arbitrary set of nodes (or states),
\item $\to$ is a binary relation on $D$ such that for all $u \in D$
  there exists $v \in D$ with $u \to v$, and
\item $\rho : D \to 2^{\Prop}$ is a labeling function that assigns to
  every node a set of atomic propositions.
\end{itemize}
A (finite relational) signature is a finite set $\sigma =
\{r_1,\ldots, r_n\}$ of relation symbols.  Every relation symbol $r
\in \sigma$ has an associated arity $\arity{r} \geq 1$.  A
$\sigma$-structure is a pair $\Struc{A} = (A,I)$, where $A$ is a
non-empty set and $I$ maps every $r \in \sigma$ to an $\arity{r}$-ary
relation over $A$.  Quite often, we will identify the relation $I(r)$
with the relation symbol $r$, and we will specify a $\sigma$-structure
as $(A, r_1, \ldots, r_n)$.

Given $\Struc{A} = (A,r_1,\ldots, r_n)$ and given a subset $B$ of $A$,
we define $r_{i \restriction B} = r_i \cap B^{\arity{r_i}}$ and
$\Struc{A}_{\restriction B} =(B, r_{1 \restriction B}, \ldots, r_{n
  \restriction B})$ (the \emph{restriction of $\Struc{A}$ to the set
  $B$}).

For a subsignature $\tau \subseteq \sigma$, a $\tau$-structure $\Bmc=
(B,J)$ and a $\sigma$-structure $\Struc{A} = (A,I)$, a
\emph{homomorphism} from $\Bmc$ to $\Struc{A}$ is a mapping $h : B \to
A$ such that for all $r \in \tau$ and all tuples
$(b_1,\ldots,b_{\arity{r}}) \in J(r)$ we have
$(h(b_1),\ldots,h(b_{\arity{r}})) \in I(r)$.  We write $\Bmc \homom
\Struc{A}$ if there is a homomorphism from $\Bmc$ to $\Struc{A}$.
Note that we do not require this homomorphism to be injective.

We now introduce \emph{constraint graphs}. These are two-sorted
structures where one part is a Kripke structure and the other part is
some $\sigma$-structure called the \emph{concrete domain}.  To connect
the concrete domain with the Kripke structure, we fix a set of unary
function symbols $\Func$. The interpretation of a function symbol from
$\Func$ is a mapping from the states of the Kripke structure to the
universe of the concrete domain.  Constraint graphs are the structures
in which we evaluate constraint $\ECTL$-formulas.

\begin{definition}
  An \emph{$\Struc{A}$-constraint graph} $\CG{C}$ is a tuple
  $(\Struc{A}, \Struc{K}, (f^{\CG{C}})_{f\in\Func})$ where:
  \begin{itemize}
  \item $\Struc{A} = (A,I)$ is a $\sigma$-structure (the concrete
    domain),
  \item $\Struc{K} = (D,\to,\rho)$ is a Kripke structure, and
  \item for each $f\in\Func$, $f^{\CG{C}}: D \to A$ is the
    interpretation of the function symbol $f$ connecting elements of
    the Kripke structure with elements of the concrete domain.
  \end{itemize}
\end{definition}

\begin{definition}
  An $\Struc{A}$-constraint path $\CG{P}$ is an $\Struc{A}$-constraint
  graph of the form $\CG{P} = (\Struc{A}, \Struc{P},
  (f^\CG{P})_{f\in\Func})$, where $\Struc{P} = (\mathbb{N}, \suc,
  \rho)$ is a Kripke structure such that $\suc$ is the successor relation
  on $\mathbb{N}$.
\end{definition}
We use $(\Struc{A}, \Struc{K}, \Func^{\CG{C}})$ as an abbreviation for
$(\Struc{A}, \Struc{K}, (f^{\CG{C}})_{f\in\Func})$.  Moreover, we often drop
the superscript $\CG{C}$ and also write constraint graph instead of
$\Struc{A}$-constraint graph if no confusion arises.

\subsection{Tree-like Structures}\label{treelike_structures}

We now introduce trees in the sense of Wolk \cite{Wolk62}, which
are also known as \emph{semi-linear orders}.  They are
partial orders $\Pmc=(P, <)$ with the additional property that for all
$p\in P$ the suborder induced by $\{p'\in P\mid p'\leq p\}$ forms a
linear order.  This property is equivalent to the one formulated by
Wolk \cite{Wolk62}: Given incomparable elements $p_1, p_2 \in P$,
there is no $q \in P$ such that $p_1 < q$ and $p_2 < q$, i.e., two
incomparable elements cannot have a common descendant.  Clearly all
trees (in the usual sense) satisfy this property, but not vice-versa.

We call a semi-linear order $\Pmc=(P,<)$ an \emph{ordinal forest}
(resp., \emph{forest}) if for all $p\in P$ the suborder induced by
$\{p'\in P\mid p'\leq p\}$ is an ordinal (resp., a finite linear
order).  A (ordinal) forest is a \emph{(ordinal) tree} if it has a unique minimal
element. A forest $\Struc{F}$ of \emph{height} $h$ (for $h\in\N$) is a
forest that contains a linear suborder with $h+1$ many elements but no
linear suborder with $h+2$ elements.  We say that an element $x \in P$
is \emph{at level} $i$ if $|\{y \in P \mid y< x\}| = i$. Thus, every minimal
element is at level $0$.

Given a partial order $(P, <)$, we denote by $\inc_<$ the
\emph{incomparability relation} defined by $p \inc_< q$ iff neither $p
\leq q$ nor $q \leq p$.  Given a $\{<,\inc, =\}$-structure $\Struc{P}
= (P,<,\inc, =)$ such that $(P,<)$ is a semi-linear order
(resp., ordinal tree, tree of height $h$), $=$ is the equality relation on $P$, and ${\inc} = {\inc_<}$, then
we also say that $\Struc{P}$ is a semi-linear order (resp. ordinal
tree, tree of height $h$).

Let us mention that the class of all countable semi-linear orders contains a 
universal structure (see Appendix \ref{sec:universalOrder}). On the
other hand, the class of all countable ordinal trees does not contain a
universal structure, but there is a fixed uncountable ordinal tree
such that all countable ordinal trees embed into this uncountable
ordinal tree.

\subsection{Logics}\label{sec-mso}

\emph{Monadic second-order logic} ($\MSO$) is the extension of
first-order logic where also quantification over subsets of the
underlying structure is allowed.  Let us fix a countably infinite set
$\VarEl$ of first-order variables that range over elements of a
structure and a countably infinite set $\VarSet$ of second-order
variables that range over subsets of a structure.  $\MSO$-formulas
over the signature $\sigma$ are given by the following grammar, where $r
\in \sigma$, $x,y, x_1,\ldots,x_{\arity{r}} \in \VarEl$, and $X \in
\VarSet$:
\begin{equation*}
  \phi ::= r(x_1,\ldots,x_{\arity{r}}) \mid x=y \mid x \in X \mid  \neg\phi
  \mid (\phi \wedge \phi) \mid \exists x  \phi \mid \exists X  \phi.
\end{equation*}
\emph{Weak monadic second-order logic} ($\WMSO$) has the same syntax
as $\MSO$ but second-order variables only range over finite subsets of
the underlying structure.

Finally, $\WMSOB$ is the extension of $\mathsf{WMSO}$ by the \emph{
  bounding quantifier} $\Bound X\, \phi$ (see \cite{BojanczykT12Long})
whose semantics is given by $\Struc{A} \models \Bound X \, \phi(X)$
(with $\Struc{A} = (A,I)$) if and only if there is a bound $b \in
\mathbb{N}$ such that $|B| \leq b$ for every finite subset $B
\subseteq A$ with $\Struc{A} \models \phi(B)$.  
The {\em quantifier rank} of a $\WMSOB$-formula is the maximal number of nested quantifiers 
(existential, universal, and bounding quantifiers) in the formula. With $\Bool(\MSO,
\WMSOB)$ we denote the set of all boolean combinations of
$\MSO$-formulas and $\WMSOB$-formulas.

\emph{Extended computation tree logic} ($\ECTL$) is a branching time
temporal logic first introduced in \cite{Thomas88,VardiW83} as an
extension of $\CTL^*$.  As the latter, $\ECTL$ is interpreted on
Kripke structures, but while $\CTL^*$ allows to specify $\LTL$
properties of infinite paths of such models, $\ECTL$ can describe
regular (i.e., $\MSO$-definable) properties of paths.  In \cite{CKL14}
we introduced an extension of $\ECTL$, called constraint $\ECTL$,
which enriches $\ECTL$ by local constraints in path formulas.

We now first recall the definition of constraint path $\MSO$-formulas, which
take the role of path formulas in constraint $\ECTL$.
\emph{Constraint path \MSO{} (over a signature $\tau$)}, denoted as
$\MSO(\tau)$, is the usual $\MSO$ for (colored) infinite paths (also
known as word structures) with a successor function $\suc$ extended by
atomic formulas that describe local constraints over the concrete
domain. Thus, given a $\tau$-structure $\Struc{A}$, $\MSO(\tau)$ can
be evaluated over the class of $\Struc{A}$-constraint paths.

In this paper we exclusively consider tree-like concrete domains over
the fixed signature $\tau =\{<,\inc, =\}$. Therefore, we simplify the
presentation and introduce constraint path $\MSO$ over $\tau$ only. For a
more general presentation we refer the reader to \cite{CKL14}.  So fix
a set $\Prop$ of atomic propositions and a set $\Func$ of unary
function symbols.  Terms and formulas of $\MSO(\tau)$ are defined by
the following grammar:
\begin{equation*}
  \begin{aligned}
    t \ \Coloneqq \ & \suc(t) \mid x \\
    \psi \ \Coloneqq \ & p(x) \mid t = t' \mid t \in X \mid \neg\psi
    \mid (\psi \wedge \psi') \mid \exists x \psi \mid \exists X \psi
    \mid f_1\suc^{i}(x) \circ f_2\suc^{j}(x)
  \end{aligned}
\end{equation*} 
where $\circ \in \tau$, $t$ and $t'$ are $\MSO(\tau)$-terms, $\psi$ and
$\psi'$ are $\MSO(\tau)$-formulas, $p \in \Prop$, $x \in \VarEl$, $X
\in \VarSet$, $i,j \in \N$ and $f_1, f_2 \in \Func$.  We call formulas of the form
$f_1\suc^{i}(x) \circ f_2\suc^{j}(x)$ for $\circ \in \tau$ \emph{atomic constraints}.  It is important to notice
that in an atomic constraint only one first-order variable $x$ is
used.

\begin{remark}
  Setting $\sigma = \{\suc\}\cup \Prop$, where $\suc$ is a unary
  function symbol and all elements of $\Prop$ are considered to be
  unary predicates, $\MSO(\tau)$ is $\MSO$ over $\sigma$ extended by
  atomic constraints over $\tau$.
\end{remark}
Let $\CG{P}= (\Struc{A}, \Struc{P}, (f^\CG{P})_{f\in\Func})$ be an
$\Struc{A}$-constraint path where $\Struc{P} = (\mathbb{N}, \suc,
\rho)$, and let $\eta: (\VarEl \cup \VarSet) \to (\mathbb{N}\cup
2^{\mathbb{N}})$ be a valuation function mapping first-order variables
to elements and second-order variables to sets. The satisfaction
relation $\models$ is defined by structural induction as follows:
\begin{align*}
  &\text{$(\CG{P}, \eta) \models p(\suc^i(x))$ iff
    $p \in \rho(\eta(x)+i)$.} \\
  &\text{$(\CG{P}, \eta) \models \suc^i(x_1) = \suc^j(x_2)$ iff
    $\eta(x_1)+i = \eta(x_2)+j$.} \\
  &\text{$(\CG{P}, \eta) \models \suc^i(x) \in X$ iff
    $\eta(x)+i \in \eta(X)$.}\\
  &\text{$(\CG{P}, \eta) \models \neg \psi$ iff it is not the case
    that
    $(\CG{P}, \eta) \models \psi$.}\\
  &\text{$(\CG{P}, \eta) \models (\psi_1 \wedge \psi_2)$ iff $(\CG{P},
    \eta) \models \psi_1$ and
    $(\CG{P}, \eta) \models \psi_2$.}\\
  &\text{$(\CG{P}, \eta) \models \exists x \psi$ iff there is an $n
    \in \N$ such that
    $(\CG{P}, \eta[x \mapsto n]) \models \psi$.}\\
  &\text{$(\CG{P}, \eta) \models \exists X \psi$ iff there is an $E
    \subseteq \N$ such that
    $(\CG{P}, \eta[X   \mapsto  E]) \models \psi$.}\\
  & \text{$(\CG{P}, \eta ) \models f_1\suc^{i}(x) \circ
    f_2\suc^{j}(x)$ iff $\Struc{A} \models f^\CG{P}_1(\rho(x)+i)
    \circ f^\CG{P}_2(\rho(x)+j) $.}
\end{align*}
For an $\MSO(\tau)$-formula $\psi$ the satisfaction relation only
depends on the variables occurring freely in $\psi$. This motivates
the following notation: If $\psi(X_1,\dots, X_m)$ is an
$\MSO(\tau)$-formula where $X_1, \dots, X_m \in \VarSet$ are the only
free variables, we write $\CG{P} \models \psi(A_1, \ldots, A_m)$ if
and only if, for every valuation function $\eta$ such that $\eta(X_i)=
A_i$, we have $(\CG{P}, \eta) \models \psi$.

Having defined  $\MSO(\tau)$-formulas  we are ready to
define constraint $\ECTL$ over the signature $\tau$ (denoted by
$\ECTL(\tau)$):

\begin{equation}\label{ECTLdef}
  \phi  \Coloneqq   
  \Ex \psi (\underbrace{\phi,\ldots, \phi}_{\text{$m$ times}}) \mid 
  (\phi \wedge \phi)  \mid \neg\phi  
\end{equation}
where $\psi (X_1, \dots, X_m)$ is an $\MSO(\tau)$-formula in which at
most the second-order variables $X_1, \ldots, X_m \in \VarSet$ are
allowed to occur freely.

$\ECTL(\tau)$-formulas are evaluated over nodes of 
$\Struc{A}$-constraint graphs.  Let $\CG{C} = (\Struc{A}, \Struc{K},
(f^{\CG{C}})_{f\in\Func})$ be an $\Struc{A}$-constraint graph, where
$\Struc{K} = (D, \to, \rho)$.  We define an infinite path $\pi$ in
$\Struc{K}$ as a mapping $\pi : \mathbb{N} \to D$ such that $\pi(i)
\to \pi(i+1)$ for all $i \geq 0$.  For an infinite path $\pi$ in
$\Struc{K}$ we define the infinite constraint path $\CG{P}_\pi =
(\Struc{A}, (\mathbb{N}, \suc, \rho'), (f^{\CG{P}_\pi })_{f\in\Func})$,
where $\rho'(n) = \rho(\pi(n))$ and $f^{\CG{P}_\pi }(n) =
f^{\CG{C}}(\pi(n))$.  Note that we may have $\pi(i) = \pi(j)$ for $i
\neq j$.  Given $d\in D$ and an $\ECTL(\tau)$-formula $\phi$, we
define $(\CG{C}, d) \models \phi$ inductively as follows:
\begin{itemize}
\item $(\CG{C}, d) \models \phi_1 \wedge \phi_2$ iff $(\CG{C}, d)
  \models \phi_1$ and $(\CG{C}, d) \models \phi_2$.
\item $(\CG{C}, d) \models \neg \phi$ iff it is not the case that
  $(\CG{C},d) \models \phi$.
\item $(\CG{C}, d) \models \Ex\psi(\phi_1, \dots , \phi_m)$ iff there
  is an infinite path $\pi$ in $\Struc{K}$ with $d = \pi(0)$ and
  $\CG{P}_\pi \models \psi(A_1, \ldots, A_m)$ where $A_i=\{j \mid j
  \geq 0, (\CG{C},\pi(j)) \models \phi_i \}$.
\end{itemize}
Note that for checking $(\CG{C}, d) \models \phi$ we may ignore all
propositions $p \in \Prop$ and all functions $f\in \Func$ that do not
occur in $\phi$.

Given a class of $\tau$-structures $\Gamma$, $\SAT(\Gamma)$ denotes
the following computational problem: \emph{ Given a formula $\phi \in
  \ECTL(\tau)$, is there a concrete domain $\Struc{A} \in \Gamma$
  and a constraint graph $\CG{C} = (\Struc{A}, \Struc{K},
  (f^{\CG{C}})_{f\in\Func})$ such that $\CG{C}\models \phi$?}  We also
write $\SAT(\Struc{A})$ instead of $\SAT(\{\Struc{A}\})$.

\subsection{Constraint $\ECTL$ and Definable Homomorphisms}

Remember that we focus our interest in this paper on the
satisfiability problem with respect to a class of structures over
the signature $\tau = \{<, \inc, =\}$ where $=$ is always interpreted as
equality and $\inc$ as the incomparability relation with respect to
$<$.  In \cite{CKL14}, we provided a connection between
$\SAT(\Struc{A})$ for some $\tau$-structure $\Struc{A}$ and the
definability of homomorphisms to $\Struc{A}$ in the logic $\Bool(\MSO,
\WMSOB)$. To be more precise, we are interested in definability of
homomorphisms to the $\{<, \inc\}$-reduct of $\Struc{A}$. In order to
facilitate the presentation of this connection, we fix a class $\Gamma$
of $\{<, \inc\}$-structures.

For every structure $\Struc{A} = (A, I) \in \Gamma$ we denote by
$\Struc{A}^=$ its expansion by equality, i.e., the $\tau$-structure
$(A,J)$ where $J(<) = I(<)$, $J(\inc) = I(\inc)$, and  $J(=) =
\{(a,a)\mid a\in A\}$.  Similarly, we set $\Gamma^= = \{\Struc{A}^=
\mid \Struc{A}\in \Gamma\}$.

We call $\Gamma^=$ \emph{negation-closed} if for every $r \in  \{<, \inc, =\}$
there is a positive existential first-order formula
$\varphi_r(x_1,\ldots,x_{\arity{r}})$ (i.e., a formula that is built
up from atomic formulas using $\wedge$, $\vee$, and $\Standardexists$)
such that for all $\Struc{A} = (A, I) \in \Gamma$
\begin{equation*}
  A^{\arity{r}}\setminus I(r) = \{ (a_1,\ldots,a_{\arity{r}}) \mid \Struc{A}
  \models \varphi_r(a_1,\ldots,a_{\arity{r}}) \}.  
\end{equation*}
In other words, the complement of every relation $I(r)$ must be
definable by a positive existential first-order formula.

\begin{example}\label{rem_neg_clos}
  For any class $\Delta$ of $\{<, \inc\}$-structures such that in every 
  $\Struc{A} \in \Delta$, (i)  $<$ is interpreted as a strict
  partial order and (ii) $\inc$ is
  interpreted as the incomparability with respect to $<$ (i.e., $x \inc y$ iff
  neither $x \leq y$ nor $y \leq x$),  $\Delta$ is negation-closed: For every $\Struc{A} \in \Delta$
  the following equalities hold in $\Struc{A}^=$, where $A$ is the universe of $\Struc{A}$:
  \begin{align*}
    (A^2 \setminus {<} ) & = \{ (x,y) \mid \Struc{A} \models
    y < x \lor y=x \lor x \inc y \}\\
    (A^2 \setminus {\inc}) & = \{ (x,y) \mid \Struc{A} \models
    x < y \vee x = y \vee y < x \}\\
    (A^2 \setminus {=} ) & = \{ (x,y) \mid \Struc{A} \models
    x < y \vee x \inc y \vee y < x \}
  \end{align*}
  In particular, the class of all semi-linear orders and all its
  subclasses are negation-closed. Note that for this it is crucial that we 
  add the incomparability relation $\inc$.
\end{example}

\begin{definition}
  We say that $\Gamma$ has the property \EHD (\emph{existence of a homomorphism  to
  a structure from $\Gamma$ is 
  $\Bool(\MSO,\WMSOB)$-definable}) if there is a
  $\Bool(\MSO,\WMSOB)$-sentence $\varphi$ such that for every
  countable $\{<, \inc\}$-structure $\Struc{B}$
  \begin{equation*}
    \Struc{B} \homom \Struc{A} \text{ for some }  \Struc{A} \in \Gamma
    \quad \Longleftrightarrow \quad  \Bmc \models \varphi. 
  \end{equation*}
\end{definition}
Now the following theorem connects $\SAT(\Gamma^=)$ with \EHD for the
class $\Gamma$.

\begin{theorem}[\cite{CKL14}] \label{thm:EHOMDef-SAT} Let $\Gamma$ be
  a class of structures over $\{<, \inc\}$.  If $\Gamma^=$ is negation-closed and $\Gamma$ has \EHD, then the problem $\SAT(\Gamma^=)$ is
  decidable.
\end{theorem}
In the next sections, we show that the following classes of tree-like
structures have \EHD:
\begin{enumerate}
\item the class of all semi-linear orders,
\item the class of all ordinal trees, and
\item for each $h\in \N$ the class of all trees of height $h$.
\end{enumerate}
Thus, Theorem~\ref{thm:EHOMDef-SAT} shows that for these classes, the satisfiability problems
for $\ECTL$ with constraints are decidable, which proves our main Theorem~\ref{thm:MainTheorem1}.

\section{Constraint $\ECTL$ over Semi-Linear Orders} \label{sec-decidable-semilinear}

Let $\Gamma$ denote the class of all  semi-linear orders
(over $\{<,\inc\}$). The aim of this section is to prove that
$\Gamma$ has \EHD.
For this purpose, we characterize
all those structures that admit a
homomorphism to some element of $\Gamma$.  
The resulting criterion can be easily translated into $\WMSO$. Hence,
we do not need the bounding quantifier from $\WMSOB$ here (the same
will be true in the following Sections~\ref{sec-ordinal-trees} and \ref{sec-bounded-height}).

It turns out that, in the case of semi-linear orders (and also ordinal
forests) the existence of such a homomorphism is in fact equivalent to
the existence of a \emph{compatible expansion}. We say that a graph\footnote{We call $(A,<,\inc)$ a graph to emphasize that
  here the binary relation symbols $<$ and $\inc$ can have arbitrary interpretations and they need
  not be a partial order and its incomparability relation. We can
  instead see them as two different kinds of edges in an arbitrary
  graph.} $(A, <, \inc)$  can be \emph{extended} to a semi-linear
order (an ordinal forest) if there is a 
partial order $\lhd$ such that $(A, \lhd, \inc_\lhd)$ is a semi-linear 
order (a ordinal forest) \emph{compatible} with $(A, <, \inc)$, i.e.,
\begin{equation}\label{compatibility}
  x < y \Rightarrow x \lhd y  \text{ and }  x \inc  y \Rightarrow x
  \inc_\lhd y.  
\end{equation}

\begin{lemma}\label{extdlpo}
   The following are equivalent for every structure $\Struc{A} = (A, <, \inc)$:
  \begin{enumerate}
  \item $\Struc{A}$ can be extended to a semi-linear order (to an ordinal forest, resp.).
  \item $\Struc{A} \homom \Struc{B}$ for some semi-linear order (ordinal tree, resp.)  $\Struc{B}$.
  \end{enumerate}
\end{lemma}

\begin{proof}
We start with the implication
($1 \Rightarrow 2$). Assume that $\Struc{A}$ can be extended to a compatible semi-linear order (ordinal forest, resp.)
$\Struc{A}' = (A, \lhd, \inc_\lhd)$. Thanks to compatibility, the identity is a homomorphism from $\Struc{A}$ to $\Struc{A}'$. In the case of an ordinal forest, one can add one common minimal element to obtain an ordinal tree.

Let us now prove
($2 \Rightarrow 1$). Suppose $h$ is a homomorphism from $\Struc{A}=(A,<,\inc)$ to some semi-linear order $\Struc{B}=(B, \prec, \inc_\prec)$. 
We extend $\Struc{A}$ to a compatible semi-linear order $(A,\lhd,\inc_\lhd)$.
Let us fix an arbitrary  well-order $<_{\text{wo}}$ on the set $A$ (which exists by the axiom of choice).
We define the binary relation $\lhd$ on $A$ as follows:
\begin{equation*}
  x \lhd y \text{ if and only if } h(x) \prec h(y) \text{ or  } (h(x) = h(y) \text{ and } x  <_{\text{wo}} y),
\end{equation*}
As usual, we denote with $\inc_\lhd$  the incomparability relation for $\lhd$, i.e., $x \inc_\lhd y$ if and only if 
neither $x \lhd y$ nor $y \lhd x$ nor $x=y$ holds.
We show that $(A, \lhd, \inc_\lhd)$ is a semi-linear order. In fact, irreflexivity and transitivity are easy consequences of the definition
of $\lhd$ and of the fact that $\prec$ is a partial order. 
To show that $\lhd$ is semi-linear, 
assume that $x_1 \lhd x$ and $x_2 \lhd x$. By definition
$h(x_1) \prec h(x)$ or $h(x_1)=h(x)$ and $h(x_2) \prec h(x)$ or $h(x_2)= h(x)$. By semi-linearity of $\Struc{B}$,  we
deduce that $h(x_1)$ and $h(x_2)$ are comparable and, by definition of $\lhd$, so are $x_1$ and $x_2$.
It remains to show that $(A,\lhd,\inc_\lhd)$ is compatible with $\Struc{A}$.
Let $x < y$. Then, by the fact that $h$ is a homomorphism, $h(x) \prec h(y)$ which guarantees that $x \lhd y$.
If $x  \inc  y$, then $h(x) \inc_{\prec} h(y)$. Since $\Struc{B}$ is a semi-linear order, this implies that neither $h(x) \prec h(y)$ nor $h(y)\prec h(x)$ nor $h(x)=h(y)$ holds.
As a consequence none of $x \lhd y$, $y \lhd x$ and $x=y$ holds. Therefore we have $x \inc_\lhd y$.

The case in which $\Struc{B}$ is an ordinal tree is dealt with similarly. It is enough to notice that $\lhd$ does not contain any infinite decreasing chains, since
$\prec$ is well-founded and $<_{\text{wo}}$ is a well-order.
\qed
\end{proof}
Inspired by Wolk's work on comparability graphs \cite{Wolk62,Wolk65}
we use Rado's selection lemma  \cite{Rado49} in order to obtain the
compactness result that a graph can be extended to a semi-linear order iff every
finite subgraph is. Recall that a choice function for family of sets $X=\{X_i \mid i \in I\}$
is a function $f$ with domain $I$ such that $f(i) \in X_i$ for all $i \in I$, i.e., it 
\emph{chooses} one element from each set $X_i$.

\begin{lemma}[Rado's selection lemma, cf.~\cite{Gottschalk51,Rado49}] \label{lemma-rado}
Let $I$ be an arbitrary index set and let $X=\{X_i \mid i \in I\}$ be a family of finite sets.
For each finite subset $A$ of $I$, let $f_A$ be a choice function for the family 
$\{X_i \mid i \in A\}$. Then there is a choice function $f$ for $X$ such that,
for all finite $A \subseteq I$, there is a finite set $B$ such that $A
\subseteq B \subseteq I$ with $f(i) = f_B(i)$ for all $i \in A$. 
\end{lemma}

\begin{lemma}[extension of Theorem~2 in \cite{Wolk65}]\label{dlpo}
A structure
$\Struc{A} = (A,<,\inc)$ can be extended to a semi-linear order if and only if
every finite substructure of $\Struc{A}$ can be
extended to a  semi-linear order. 
\end{lemma}

\begin{proof}
The direction $(\Rightarrow)$ is trivial.
For the direction $(\Leftarrow)$, let
$$I = \{\{x,y\} \subseteq A \mid x \neq y\}$$ be the set of pairs of
distinct elements of $A$. 
For all $i = \{x,y\} \in I$ we define $Z_i = \{ (x,y), (y,x), \# \}$. 
We want to find a choice function for the family of sets $\{Z_i \mid i \in I\}$ which is
in some sense \emph{compatible} with the relations $\inc$ and $<$. In fact, choosing for each $i \in I$ one element of $Z_i$ corresponds intuitively to deciding whether the two elements  $x$ and $y$ are comparable, and in which order, or if they are incomparable.

Each finite subset $J$ of $I$ defines a 
set $\bar J =  \{x \in j \mid j \in J \}$ and a 
substructure $\Struc{A}_{\restriction{\bar J}} = (A_{\restriction{\bar J} },
<_{\restriction{\bar J}}, \inc_{\restriction{\bar J}})$. 
Since $\Struc{A}_{\restriction{\bar J}}$ is finite, by hypothesis it can be
extended to a semi-linear order. Hence, we can find a partial order $\lhd_J$ on $\bar J$
such that $(A_{\restriction{\bar J}},\lhd_{ J}, \inc_{\lhd_J})$ is a semi-linear order compatible with
$\Struc{A}_{\restriction{\bar J}}$ as in
\eqref{compatibility} on page \pageref{compatibility}. 

Let $f_J$ be the choice function for $\{ Z_j \mid j \in J\}$ defined as follows:
\begin{equation}\nonumber
f_J (\{x,y\}) = \left\{ 
\begin{array}{cl}
(y,x) & \text{ iff } y \lhd_J x,\\
(x,y) & \text{ iff } x \lhd_J y,\\
\# & \text{ otherwise}.
\end{array} \right.
\end{equation}
By Lemma~\ref{lemma-rado} we can find a choice function $f$ for $\{ Z_i \mid i
\in I\}$ such that for all finite $J \subseteq I$  there is a finite
set $K$ such that 
\begin{equation*}
  J \subseteq K \subseteq I  \text{ and } f(j) = f_K(j)  \text{ for all } j \in J.
\end{equation*}
Define $x\lhd y$ iff $(x,y) \in f(I)$.
We need to prove that $(A,\lhd, \inc_\lhd)$ is an extension of $\Struc{A}$ to a semi-linear order.
But all the properties that we need to check are local, and thanks to Rado's selection lemma, 
$\lhd$ always coincides, on every finite subset of $A$, with some
$\lhd_J$, which is a semi-linear order compatible with $<$ and $\inc$. 
\qed
\end{proof}
Thanks to Lemma~\ref{dlpo}, given a $\{<, \inc\}$-structure
$\Struc{A}$, proving \EHD only requires
to look for a necessary and sufficient condition which
guarantees that every finite substructure of $\Struc{A}$
admits a homomorphism into a semi-linear order.

\begin{definition}\label{def:central_point}
Let $\Struc{A} = (A,<,\inc)$ be a graph. Given $A' \subseteq A$, we say $A'$ is
\emph{connected (with respect to $<$)} if and only if, for all $a,a'\in
A'$ , there are $a_1, \dots, a_n \in A'$ such that $a = a_1 $, $a'=
a_n$ and $a_i < a_{i+1}$ or $a_{i+1} < a_i$ for all $1 \leq i \leq n-1$.
A {\em connected component} of $\Struc{A}$  is a maximal (with respect to inclusion) connected 
subset of $A$.
Given a subset $A' \subseteq A$  and $c \in A'$, we say that $c$ is
a \emph{central point of $A'$} if and only if for every  $a \in A'$ neither
$a  \inc  c$ nor $c  \inc  a$ nor $a<c$ holds.
\end{definition}
In other words,   a
central point of  a subset $A' \subseteq A$  
is a node 
of the structure $\Struc{A} = (A,<,\inc)$
which has no incoming or outgoing $\inc$-edges, and no incoming
$<$-edges within $A'$.  

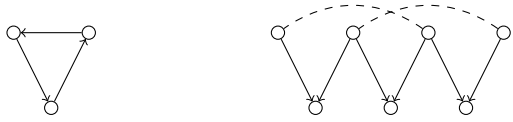
\begin{figure}
  \centering
  
  \begin{tikzpicture}
    [rotate = 0, transform shape, scale=1 ]   
    \begin{scope}
      [
      node distance = 1 and 0.5,
      on grid,
      basickonf/.style={ 
        circle, draw, solid, thin, minimum size = 5, inner sep =0
      },
      konf/.style={ basickonf },
      less/.style= { <- },
      greater/.style= { -> },
      inc/.style= { dashed, bend right=35 },
      ]
      \begin{scope}
        \node[konf] (left) {};
        \node[konf, below right = of left] (down) {} 
        edge[less] (left)
        ;
        \node[konf, above right = of down] {} 
        edge[less] (down)
        edge[greater] (left)
        ;
        
      \end{scope}
      
      \begin{scope}[xshift = 100]

        
        \foreach \pref in {0, ..., 0}
        \foreach \row in {0, ..., 0}
        {
          \node [konf,] (\pref-\row-l) at ($(5*\pref, 4*\row)$) {};
          \node [konf, below right = of \pref-\row-l] (\pref-\row-b_1)    {}
          edge [less] (\pref-\row-l)
          ;
          \node [konf,  above right = of \pref-\row-b_1] (\pref-\row-a_0)    {}
          edge [greater] (\pref-\row-b_1)
          ;
          \node [konf, below right  = of \pref-\row-a_0] (\pref-\row-b_2)    {}
          edge [less] (\pref-\row-a_0)
          ;
          
          \node [konf, above right = of \pref-\row-b_2] (\pref-\row-c_0)    {}
          edge [greater] (\pref-\row-b_2)
          edge[inc] (\pref-\row-l)
          ;
          \node [konf, below right  = of \pref-\row-c_0] (\pref-\row-b_3)    {}
          edge [less] (\pref-\row-c_0)
          ;
          
          \node [konf, above right = of \pref-\row-b_3] (\pref-\row-r)    {}
          edge [greater] (\pref-\row-b_3)
          edge [inc] (\pref-\row-a_0)
          ;

        }
      \end{scope}
    \end{scope}
  \end{tikzpicture}
 \caption{\label{cycle} A $<$-cycle of three elements and an
   ``incomparable triple-u'', where dashed lines are $\inc$-edges.}
\end{figure}

\begin{example}
A $<$-cycle (of any number of elements) does not have a central point,
nor does an \emph{incomparable triple-u}, see Figure~\ref{cycle}. 
Both structures do not admit homomorphism into a semi-linear order. While this statement is obvious for the cycle, we leave the proof for the incomparable triple-u as an exercise.
\end{example}

\begin{lemma}\label{compactness}
A finite structure $\Struc{A} = (A, <, \inc)$ can be extended to a semi-linear order if and only if
every non-empty connected $B\subseteq A$ has a central point.  
\end{lemma}
Let us  extract the main argument for the
$(\Rightarrow)$-part of the proof for later reuse:

\begin{lemma}\label{lem:minIsCentral}
  Let $(A, \lhd, \inc_\lhd)$ be a semi-linear order extending $\Struc{A} = (A, <, \inc)$. If
  a connected subset $B\subseteq A$ (with respect to $<$) contains a minimal element $m$ with respect to
  $\lhd$, then $m$ is central in $B$ (again with respect to $\Struc{A}$). 
\end{lemma}
\begin{proof}
  Let $b \in B$. Since $B$ is connected, there are
  $b_1, \ldots, b_n \in B$ such that 
  $b_1 = m$, $b_n = b$ and 
  $b_{i} < b_{i+1}$ or $b_{i+1} < b_{i}$ for all $ 1\leq i \leq n-1$.
  As $\lhd$ is compatible with $<$, this implies that $b_{i} \lhd b_{i+1}$ or 
  $b_{i+1} \lhd b_{i}$ for all  $1 \leq i \leq  n-1$.
  Given that $m$ is minimal, applying semi-linearity of $\lhd$, we
  obtain that $m = b_i$ or $m \lhd b_i$ for all $1 \leq i \leq n$.
  In particular, we have $m = b$ or $m \lhd b$.
  Since $(A, \lhd, \inc_\lhd)$ is a semi-linear order, compatible with $(A, <, \inc)$,
  we cannot have  $b < m$, $m  \inc  b$ or $b  \inc  m$ (since this would imply
  $b \lhd m$ or $m  \inc_\lhd  b$).  Hence, $m$ is central.
  \qed
\end{proof}
\begin{proof}[of Lemma~\ref{compactness}]
For the direction
($\Rightarrow$) let $B$ be any non-empty connected subset of $A$. 
Since $B$ is finite, there is a minimal element $m$. Using the 
previous lemma we conclude that $m$ is central in $B$. 

We prove the direction 
($\Leftarrow$)
by induction on $n=|A|$. Suppose $n=1$ and let $A=\{a\}$. 
The fact that $\{a\}$ has a central point implies that neither $a<a$
nor $a  \inc  a$ holds. Hence, $\Struc{A}$ is a semi-linear order.

Suppose $n > 1$ and assume the statement to be true for all $i < n$.
If $\Struc{A}$ is not connected with respect to $<$, then we apply the induction hypothesis to every
connected component. The union of the resulting semi-linear
orders extends $\Struc{A}$. 
Now assume that $\Struc{A}$ is connected and
let $c$ be a central point of $A$.
By the  inductive hypothesis we can find 
$\lhd'$ such that $(A\setminus\{c\}, \lhd', \inc_{\lhd'})$ is a semi-linear order 
extending $\Struc{A}\setminus\{c\}$. 
We define $\lhd := \lhd' \cup \{(c,x) \mid x \in A\setminus \{c\}\}$
(i.e., we add $c$ as a smallest element),
which is obviously a partial order on $A$. 

To prove that $\lhd$ is semi-linear, let $a_1,a_2,a \in A$ such that
$a_1 \lhd a$ and $a_2 \lhd a$.
If $a_1=c $ or $a_2=c$, then $a_1$ and $a_2$ are comparable by
definition. Otherwise, we conclude that $a_1,a_2,a \in A
\setminus\{c\}$. Hence, $a_1 \lhd' a$ and $a_2 \lhd' a$, and semi-linearity of $\lhd'$ settles the claim. 

We finally show compatibility.
Suppose that $a < b$.
If $a = c$, then $a \lhd b$.  The case $b = c$ cannot
occur, because $c$ is central in $A$. 
The remaining possibility  $a\neq c \neq b$ implies that 
$a \lhd' b$  and hence $a\lhd b$ as desired.
Finally, suppose that $a  \inc  b$. Then $a \neq c \neq b$, because $c$ is
central. We conclude that $a \inc_{\lhd'} b$ and also
$a\inc_\lhd b$. 
\qed
\end{proof}

We are finally ready to state the main result of this section which
(together with Theorem \ref{thm:EHOMDef-SAT}) completes the proof of
the first part of Theorem \ref{thm:MainTheorem1}:
\begin{proposition}\label{prop:EHomDefDLPO}
The class of all semi-linear orders $\Gamma$ has  \EHD.
\end{proposition}
\begin{proof}
  Take $\Struc{A} = (A, <, \inc)$. 
  Thanks to Lemmas
  \ref{extdlpo}, \ref{dlpo} and \ref{compactness},
  it is enough to show that $\WMSO$ can
  express the condition that every finite and non-empty connected
  substructure of $\Struc{A}$ has a central point. 
  We define the following $\WMSO$-formula $\reach(x,y,X)$ such that
  $\Struc{A}\models \reach(a,b,B) $ if and only if $a$ and $b$ are in the same connected component of $\Struc{A}_{\restriction{B}}$:
\begin{equation*}
x \in X \land 
     \forall{Y \subseteq X} \big[  \big( x\in Y \land \forall{z \in Y} \, \forall{w \in X}
     (\phi(z,w) \rightarrow w \in Y )\big)
       \rightarrow y \in Y \big] ,
 \end{equation*}
where $\phi(z,w) := z<w \vee w<z$.
Then, we define the following $\WMSO$-formulas:
  \begin{align*}
    \mathsf{connected}(X) &:= \forall{x \in X} \, \forall{y \in X}  \, \reach(x,y,X),\\
    \mathsf{central}(x,X) &:= x \in X \wedge \forall{y \in X} \,
    \neg(x  \inc  y \vee y  \inc  x \vee y < x),\text{ and}\\
    \psi &:= \forall{X} (\mathsf{connected}(X) \wedge X \neq \emptyset \rightarrow 
    \exists{x}  \mathsf{central}(x,X) ).  
  \end{align*}
  It is straightforward to verify that $\Struc{A} \models \psi$ if and only
  if every finite non-empty connected subset of $A$ has a central point. 
\qed
\end{proof}

\section{Constraint $\ECTL$ over Ordinal Trees} \label{sec-ordinal-trees}

Let $\Omega$ denote the class of all ordinal trees
(over the signature $\{<,\inc\}$). The aim of this section is to prove that
 $\Omega$ has \EHD as well. 
We use again the notions of a connected subset and a central point as
introduced in Definition~\ref{def:central_point}. We will
characterize those structures which admit a homomorphism into an
ordinal tree.  Here, in contrast with the case of semi-linear orders,
the final condition will be that all connected sets (not just the
finite ones) have a central point.

\begin{lemma}\label{lem_ord_trees}
Let $\Struc{A} =(A, <, \inc)$ be a structure. There exists $\Struc{O} \in \Omega$ such that $\Struc{A}
\homom \Struc{O}$ if and only if every non-empty and connected $B
\subseteq A$ has a central point. 
\end{lemma}

\begin{proof}
  We start with the direction ($\Rightarrow$).  Due to Lemma
  \ref{extdlpo} we can assume that there is a relation $\lhd$ that
  extends $(A, <, \inc)$ to an ordinal forest.  Let $B \subseteq A$ be
  a non-empty connected set.  Since $(A, \lhd, \inc_\lhd)$ is an
  ordinal forest, $B$ has a minimal element $c$ with respect to
  $\lhd$. By Lemma \ref{lem:minIsCentral}, $c$ is a central point of
  $B$.

  For the direction ($\Leftarrow$) we first define a partition of the
  domain of $\Struc{A}$ into subsets $C_\beta$ for $\beta \sqsubset
  \chi$, where $\chi$ is an ordinal (whose cardinality is bounded by
  the cardinality of $A$). Here $\sqsubset$ denotes the
  natural order on ordinals.  Assume that the pairwise disjoint
  subsets $C_\beta$ have been defined for all $\beta \sqsubset \alpha$
  (which is true for $\alpha=0$ in the beginning).  Then we define
  $C_\alpha$ as follows. Let $C_{\sqsubset\alpha} = \bigcup_{\beta
    \sqsubset \alpha} C_\beta \subseteq A$.  If $A \setminus
  C_{\sqsubset\alpha}$ is not empty, then we define
  $\mathcal{C}_\alpha$ as the set of connected components of $A
  \setminus C_{\sqsubset\alpha}$.  Let
$$
C_\alpha = \{ c \in A \setminus C_{\sqsubset\alpha} \mid c \text{ is a
  central point of some } B \in \mathcal{C}_\alpha \} .
$$
Clearly, $C_\alpha$ is not empty. Hence, there must exist a smallest
ordinal $\chi$  such that $A = C_{\sqsubset\chi}$.

For every ordinal $\alpha\sqsubset\chi$ and each element $c \in
C_\alpha$ we define the sequence of connected components
$\road(c)=(B_\beta)_{(\beta \sqsubseteq \alpha)}$, where $B_\beta \in
\mathcal{C}_\beta$ is the unique connected component with $c \in
B_\beta$.  This ordinal-indexed sequence keeps record of the
\emph{road} we took to reach $c$ by storing information about the
connected components to which $c$ belongs at each stage of our
process.

Given $\road(c) = (B_\beta)_{(\beta \sqsubseteq \alpha)}$ and
$\road(c')=(B'_\beta)_{(\beta \sqsubseteq \alpha')}$ for some $c \in
C_{\alpha}$ and $c' \in C_{\alpha'}$, let us define $\road(c) \lhd
\road(c')$ if and only if $\alpha \sqsubset \alpha'$ and
$B_\beta=B'_\beta$ for all $\beta \sqsubseteq \alpha$. Basically this
is the \emph{prefix order} for ordinal-sized sequences of connected
components.

Now let $O=\{\road(c) \mid c \in A\}$. Note that $\Struc{O} = (O,
\lhd, \inc_\lhd)$ is an ordinal forest, because for each $c\in
C_\alpha$ the order $(\{\road(c') \mid \road(c') \unlhd \road(c)\},
\unlhd)$ forms the ordinal $\alpha$ (for each $\beta\sqsubset\alpha$
it contains exactly one road of length $\beta$).

Now we show that the mapping $h$ with $h(c) = \road(c)$ is a
homomorphism from $\Struc{A}$ to $\Struc{O}$.  Take elements $a,a' \in
A$ with $a \in C_{\alpha}$, and $a' \in C_{\alpha'}$ for some $\alpha,
\alpha' \sqsubset \chi$. Let $\road(a)=(B_\beta)_{(\beta \sqsubseteq
  \alpha)}$ and $\road(a')=(B'_\beta)_{(\beta \sqsubseteq \alpha')}$.
\begin{itemize}
\item If $a<a'$, then (i) $\alpha \sqsubset \alpha'$, because $a'$
  cannot be central point of a set which contains $a$, and (ii)
  $B_\beta = B'_\beta$ for all $\beta \sqsubseteq \alpha$ because $a$
  and $a'$ belong to the same connected component of $A \setminus
  C_{\sqsubset\beta}$ for all $\beta \sqsubseteq \alpha$.  By these
  observations we deduce that $\road(a)\lhd \road(a')$.
\item If $a \inc a'$, then, without loss of generality, suppose that
  $\alpha \sqsubseteq \alpha'$.  At stage $\alpha$, $a$ is a central
  point of $B_\alpha \in \mathcal{C}_{\alpha}$. Since $\alpha
  \sqsubseteq \alpha'$, the connected component $B'_\alpha$ exists.
  We must have $B_\alpha \neq B'_\alpha$, since otherwise we would
  have $a \inc a' \in B_\alpha$ contradicting the fact that $a$ is
  central for $B_\alpha$.  Therefore, $\road(a) \inc_\lhd \road(a')$.
\end{itemize}
We finally add one extra element $\road_0$ and make this the minimal
element of $\Struc{O}$, thus finding a homomorphism from $\Struc{A}$
into an ordinal tree.  \qed
\end{proof}
We can now complete the proof of the second part of Theorem
\ref{thm:MainTheorem1} 
\begin{proposition}\label{prop:EHomDefOrdinalTree}
The class $\Omega$ of all  ordinal trees has  \EHD.
\end{proposition}

\begin{proof}
  Given a $\{<,\inc\}$-structure $\Struc{A}$, it suffices by Lemma~\ref{lem_ord_trees} to find an
  $\MSO$-formula expressing the fact that every non-empty connected
  subset of $\Struc{A}$ has a central point. 
  Recall the $\WMSO$-formula $\psi$ from Theorem~\ref{prop:EHomDefDLPO}. Seen as an $\MSO$-formula, $\psi$ clearly
  does the job.
  \qed
\end{proof}

\begin{remark}
  \label{rem-ordinal-tree-proof}
  
  The procedure described in the proof of Lemma~\ref{lem_ord_trees}
  can be also used to embed a structure $\Struc{A} =(A, <, \inc)$ into
  an ordinary tree (where for every $x$, the set of all elements
  smaller than $x$ forms a finite linear order). For this, the ordinal
  $\chi$ has to satisfy $\chi \leq \omega$, i.e., every element $a \in
  A$ has to belong to a set $C_n$ for some finite $n$. We use
  this observation in Section~\ref{sec-trees}. Unfortunately, 
  our results from Section~\ref{sec-trees} imply that 
  $\chi \leq \omega$ cannot be expressed in $\Bool(\MSO,\WMSOB)$.
\end{remark}

\section{Constraint $\ECTL$ over Trees of Height
  $h$} \label{sec-bounded-height} 

Fix $h\in\N$.  The aim of this section is to show that the class
$\Theta_h$ of all trees of height $h$ (over $\{<, \inc\}$) has \EHD.
The proof relies on the fact that we can unfold the fixpoint procedure
on the central points from the ordinal tree setting for $h$ steps in
$\MSO$. 
  
For this section, we fix an arbitrary structure 
$\Struc{A} = (A, <,\inc)$.  We first define subsets 
$A_0, A_1, \dots, A_h\subseteq A$ 
that are pairwise disjoint.  The elements of $A_0$ are the central
points of $A$ (this set is possibly empty) and, for each $i\geq 1$, $A_i$ contains the central
points of each connected component of $A\setminus (A_0 \cup\dots\cup
A_{i-1})$. Note that $A_0$ contains exactly those nodes of $\Struc{A}$
that a homomorphism from $\Struc{A}$ to some tree can map to the root
of the tree because elements from $A_0$ are neither incomparable to any other element nor
below any other element, while all element outside of $A_0$ have to be
incomparable to some other element or have to be below some other element.
Hence they cannot be mapped to the root by any homomorphism. 
Thus, there is a homomorphism from $\Struc{A}$  to some element of
$\Theta_h$ if and only if $\Struc{A}\setminus A_0$ can be embedded
into some forest of height $h-1$. Now the sets $A_i$ for $1\leq i \leq
h$ collect exactly those elements which are chosen in the $i$-th step
of the fixpoint procedure from the proof of
Lemma~\ref{lem_ord_trees} (where this set is called $C_i$). Thus, if $A_0, A_1,
\dots, A_h$ form a partition of $A$, then $\Struc{A}$ allows a
homomorphism to some $\Struc{T}\in\Theta_h$. 
It turns out that the converse is also true. 
If $\Struc{A} \homom \Struc{T}$ for some
$\Struc{T} \in \Theta_h$  then $A_0, A_1, \dots, A_h$ form a partition of
$A$.  Thus, it suffices to show that each $A_i$ is $\MSO$-definable.
To do this, we define for all $i\in\N$ the formulas
  \begin{eqnarray*}
    \inA{0}(x) &:=& \forall y \neg (y < x \vee 
    y\inc x \vee x \inc y) ,\\    
    \inA{i+1}(x) &:=& \forall y (  \mathsf{con}_{i+1}(x,y) \rightarrow
    \neg (y < x \vee y\inc x \vee x \inc y)), \\
    \mathsf{con}_{i+1}(x,y) & := & \exists Z \forall z (z\in Z \rightarrow
    \bigwedge_{j=0}^{i} \neg \inA{j}(z)) \land \reach(x,y,Z),
  \end{eqnarray*}
  where $\reach(x,z,Z)$ is defined as in the proof of Proposition~\ref{prop:EHomDefDLPO} on page \pageref{prop:EHomDefDLPO}.
  Let $A_0$ be the set of nodes $a\in A$ such that
  $\Struc{A}\models\inA{0}(a)$ and
  let $A_{i+1}$
  be the set of nodes $a\in
  A$ such that $\Struc{A}\models\inA{i+1}(a)$. 

Clearly $A_0$ is the set of central points of $A$. 
Inductively, one shows that
$A_{i+1}$ is the set of central points of the connected components of
$A\setminus (A_0 \cup\dots\cup A_i)$. 

\begin{lemma} \label{lemma-trees-height-k}
  There exists $\Struc{T} \in \Theta_h$ such that $\Struc{A} \homom \Struc{T}$
  if and only if $A_0, A_1, \dots, A_h$ is a partition of $A$. 
\end{lemma}

\begin{proof}
For the direction $(\Rightarrow)$ take a homomorphism $g$ from $\Struc{A}$
to a tree $\Struc{T} = (T, \lhd, \inc_\lhd) \in \Theta_h$. By induction we prove that if
  $g$ maps $a$ to the $i$-th level of $\Struc{T}$ then $a\in A_j$ for some $j\leq
  i$.  For $i = 0$ assume that $g(a)$ is the root of the tree. Then $a$
  cannot be incomparable or greater than any other element. Thus, it
  is a central point of $A$, i.e., $a\in A_0$.

  For the inductive step, assume that $g(a)$ is on the $i$-th level for $i >0$.
  Heading for a contradiction, assume that $a$ is 
  neither in $A_0\cup\dots\cup A_{i-1}$ nor a
  central point of
  some connected component of $A\setminus (A_0 \cup\dots\cup
  A_{i-1})$. Then there is some 
  $a'\in A\setminus (A_0 \cup \dots \cup A_{i-1})$ such that $a$ and
  $a'$ are in the same connected 
  component of $A\setminus (A_0\cup\dots\cup A_{i-1})$
  and one of $a' < a$, $a' \inc a$ or $a \inc a'$ holds. Since $g$ is 
  a homomorphism, we get
  $g(a') \lhd g(a)$ or $g(a') \inc_\lhd g(a)$. 
  If $g(a') \lhd g(a)$, then  $a'$ has to be mapped by $g$ to some level $j<i$,
  whence $a'\in A_0\cup \dots \cup A_j$ by the inductive hypothesis. This
  contradicts our assumption on $a'$. Now, assume that
  $g(a') \inc_\lhd g(a)$. Let $a=a_0, a_1, \ldots, a_m=a'$ be a
  path connecting $a$ and $a'$ in 
  $A\setminus (A_0\cup\dots\cup A_{i-1})$.  Since  $a_i\notin A_0\cup \dots \cup A_{i-1}$,
  the inductive hypothesis shows that all $g(a_i)$ are on level $i$ or larger.
  But then, since  $a_0, a_1, \ldots, a_m$ is a path, all 
  $g(a_i)$ must belong to the subtree rooted at  $g(a)$. This leads to the
  contradictions that $g(a_m) = g(a')$ is in the subtree rooted at
  $g(a)$ and hence is not incomparable to $g(a)$. Thus, we can conclude that $a\in A_0\cup \dots \cup A_i$. 
  
  For the direction  $(\Leftarrow)$ assume that $A_0\cup \dots \cup A_h = A$. 
  Applying the same construction 
  described in the proof of Lemma~\ref{lem_ord_trees} for
  ordinal trees, it is not hard to see that we find a homomorphism
  $g$ from $\Struc{A}$ to some tree of height $h$
  which maps the elements of
  $A_i$ to elements on level $i$. 
  Should $A_0$ be empty, then $A$ would not be connected, and we would
  have a forest of height $h-1$. Adding a minimal element we still get
  a tree of height $h$.
  \qed  
\end{proof}

\begin{theorem}
  $\Theta_h$ has  \EHD.
\end{theorem}
\begin{proof}
  Let $\Struc{A}$ be any $\{<,\inc\}$-structure. Then, by Lemma~\ref{lemma-trees-height-k},
  $\Struc{A} \homom \Struc{T}$ for some $\Struc{T} \in \Theta_h$ if and
  only if 
\begin{equation*}
\Struc{A}\models \forall x \bigvee_{i=0}^h \inA{i}(x) .
\end{equation*}
\end{proof}

\section{Trees do not have  \EHD} \label{sec-trees}

Let $\Theta$ be the class of all countable trees (over  $\{<,\inc\}$).
In this section, we prove that the logic $\Bool(\MSO,\WMSOB)$ 
(the most expressive logic for which the \EHD-technique currently works) cannot distinguish between
graphs that admit a homomorphism to some element of $\Theta$ and those
that do not.   Heading for a contradiction, assume that $\varphi$ is a sentence such
that a countable structure $\Struc{A} =(A,<,\inc)$ 
satisfies $\varphi$  if and only if there is a homomorphism from
$\Struc{A}$ to some $\Struc{T} \in \Theta$.  
Let $k$ be the quantifier rank of $\varphi$. We construct two
graphs  $\Struc{E}_k$ and
$\Struc{U}_k$ such that $\Struc{E}_k$ admits a homomorphism into a
tree while $\Struc{U}_k$ does not. We then use an Ehrenfeucht-\Fraisse game
for $\Bool(\MSO,\WMSOB)$
to show that $\varphi$ cannot separate these two structures,
contradicting our assumption. This contradiction shows that $\Theta$
does not have $\EHD$ proving our second
main result Theorem~\ref{thm:MainTheorem2}.

\subsection{The $\WMSOB$-Ehrenfeucht-\Fraisse-game}
\label{sec:EFGame}

The $k$-round $\WMSOB$-EF-game on a pair of structures $(\Struc{A},
\Struc{B})$ over the same finite relational signature $\sigma$ is
played by spoiler and duplicator as follows.\footnote{For the ease of
  presentation we assume that
  $\Struc{A}$ and $\Struc{B}$ are infinite structures.}
In the following, $A$ denotes the domain of $\Struc{A}$ and $B$ the
domain of $\Struc{B}$. 

The game starts in position 
\begin{equation*}
  p_0:=(\Struc{A}, \emptyset, \emptyset, \Struc{B}, \emptyset, \emptyset).  
\end{equation*}
In general, before playing the $i$-th round (for $1\leq i \leq k$) 
the game is in a position 
\begin{equation*}
  p = (\Struc{A}, a_1, a_2, \dots, a_{i_1}, A_1, A_2, \dots, A_{i_2},
       \Struc{B}, b_1, b_2, \dots, b_{i_1}, B_1, B_2, \dots, B_{i_2}),
\end{equation*}
where  
\begin{enumerate}
\item $i_1,i_2\in\N$ satisfy $i_1+i_2=i-1$, 
\item $a_j\in A$ for all $1\leq j \leq i_1$,
\item $b_j\in B$ for all $1\leq j \leq i_1$,
\item $A_j\subseteq A$ is a finite set for all $1\leq j \leq i_2$, and
\item $B_j\subseteq B$ is a finite set for all $1\leq j \leq i_2$.
\end{enumerate}
In the $i$-th round spoiler and duplicator produce the next
position as follows. Spoiler chooses to play one of the following
three possibilities:
\begin{enumerate}
\item Spoiler can play an \emph{element move}. For this he chooses
  either some $a_{i_1+1}\in A$ or $b_{i_1+1}\in B$. 
  Duplicator then responds with an element from the other structure,
  i.e., with $b_{i_1+1}\in B$ or $a_{i_1+1}\in A$. 
  The position in the next round is
\begin{equation*}
  (\Struc{A}, a_1, a_2, \dots, a_{i_1+1}, A_1, A_2, \dots, A_{i_2},
       \Struc{B}, b_1, b_2, \dots, b_{i_1+1}, B_1, B_2, \dots, B_{i_2}).
\end{equation*}
\item 
  Spoiler can play a \emph{set move}. For this he chooses
  either some finite $A_{i_2+1}\subseteq A$ or some finite $B_{i_2+1}\subseteq B$. 
  Duplicator then responds with a finite set from the other structure,
  i.e., with $B_{i_2+1}\subseteq B$ or $A_{i_2+1}\subseteq A$. 
  The position in the next round is
  \begin{equation*}
    (\Struc{A}, a_1, a_2, \dots, a_{i_1}, A_1, A_2, \dots, A_{i_2+1},
    \Struc{B}, b_1, b_2, \dots, b_{i_1}, B_1, B_2, \dots, B_{i_2+1}).
  \end{equation*}
\item 
  Spoiler can play a \emph{bound move}. 
  For this he chooses
  one of the structures $\Struc{A}$ or $\Struc{B}$ and chooses a
  natural number $l \in \N$.
  Duplicator responds with another number $m \in \N$. 
  Then the game continues as in the case of a set move with the
  restrictions that spoiler has to choose a subset of size at least
  $m$ from his chosen structure and duplicator has to respond with a
  set of size at least $l$. 
\end{enumerate}
After $k$ rounds, the game ends in a position 
\begin{equation*}
  p = (\Struc{A}, a_1, a_2, \dots, a_{i_1}, A_1, A_2, \dots, A_{i_2},
       \Struc{B}, b_1, b_2, \dots, b_{i_1}, B_1, B_2, \dots, B_{i_2}).
\end{equation*}
Duplicator wins the game if 
\begin{enumerate}
\item  $a_j\in A_k \Leftrightarrow b_j\in B_k$ for all $1\leq j \leq
  i_1$ and all $1\leq k \leq i_2$,
\item $a_j = a_k \Leftrightarrow b_j = b_k$ for all $1\leq j<k \leq
  i_1$, and
\item for all relation symbols $R\in\sigma$ (of arity $n$)
  $(a_{j_1}, a_{j_2}, \dots, a_{j_n})\in R^{\Struc{A}} \Leftrightarrow
  (b_{j_1}, b_{j_2}, \dots, b_{j_n})\in R^{\Struc{B}}$ for all
  $j_1, j_2, \ldots, j_n \in \{1, \ldots, i_1\}$. 
\end{enumerate}
As one would expect, the $\WMSOB$-EF-game can be used to show
undefinability results for $\WMSOB$ due to the relationship between
winning strategies in the $k$-round game and equivalence with respect
to formulas up to quantifier rank $k$.

\begin{proposition}
  For given $\sigma$-structures $\Struc{A}$ and $\Struc{B}$, elements
  $a_1, a_2, \dots, a_{i_1}\in\Struc{A}$, $b_1,b_2, \dots,
  b_{i_1}\in\Struc{B}$ and finite sets $A_1, A_2, \dots,
  A_{i_2}\subseteq\Struc{A}$, $B_1, B_2, \dots, B_{i_2}\subseteq\Struc{B}$, 
  define the position 
  \begin{equation*}
    p=(\Struc{A}, a_1, \dots, a_{i_1}, A_1, \dots, A_{i_2}, \Struc{B}, b_1,
    \dots, b_{i_1}, B_1, \dots, B_{i_2}).  
  \end{equation*}
  Then, 
  $(\Struc{A}, a_1, a_2, \dots, a_{i_1}, A_1, A_2, \dots, A_{i_2})$ and
  $(\Struc{B}, b_1, b_2, \dots, b_{i_1}, B_1, B_2, \dots, B_{i_2})$ are
  indistinguishable by any $\WMSOB$-formula 
  $\varphi(x_1, \dots, x_{i_1}, X_1, \dots, X_{i_2})$ of quantifier rank $k$
  if and only if duplicator has a winning strategy in the $k$-round
  $\WMSOB$-EF-game started in $p$. 
\end{proposition}

\begin{proof}
  First of all note that up to logical equivalence there are only finitely many different $\WMSOB$-formulas 
  $\varphi(x_1, \dots, x_{i_1}, X_1, \dots, X_{i_2})$ of
  quantifier rank $k$. This fact is
  proved in a completely analogous way to the case of first-order or 
  monadic second-order logic. 

  The proof is by induction on $k$. The base case $k=0$ is trivial. 
  Assume now that the proposition holds for $k-1$. We use the abbreviations $\bar a = (a_1, \dots, a_{i_1})$,
      $\bar A = (A_1, \dots, A_{i_2})$, $\bar b = (b_1, \dots, b_{i_1})$, and
      $\bar B = (B_1, \dots, B_{i_2})$  in the following.
 First assume that there is a 
    \WMSOB-formula 
     $\varphi(x_1, \dots, x_{i_1}, X_1, \dots, X_{i_2})$
    of quantifier rank $k$ such that
    \begin{equation}
      \label{eq:sat}
      \Struc{A} \models \varphi(\bar a, \bar A)  
    \end{equation}
    and
    \begin{equation}
      \label{eq:nonsat}
      \Struc{B} \not\models \varphi(\bar b, \bar B).      
    \end{equation}
    We show that spoiler has a winning strategy in the $k$-round game by
    a case distinction on the structure of $\varphi$. 
    We only consider the case $\varphi = \Bound X \psi$
    (all other cases can be handled exactly as in the $\WMSO$-EF-game,
    see e.g.~\cite{EbbinghausF95}).
    Let $l\in\N$ be a strict bound witnessing 
    \eqref{eq:sat}, in the sense that there is no set $A_{i_2+1}$ of size at least
    $l$ such that $\Struc{A} \models \varphi(\bar a, \bar A, A_{i_2+1})$.
     Then spoiler chooses structure $\Struc{B}$ and
    bound $l$.  Duplicator responds with some bound $m\in\N$. 
    Due to \eqref{eq:nonsat} 
    \begin{equation*}
      \Struc{B} \models
      \neg \Bound X \psi(\bar b, \bar B,X) .
    \end{equation*}
    Hence,
    there is a set $B_{i_2+1}$ of size at least $m$ such that
    \begin{equation*}
      \Struc{B} \models \psi(\bar b, \bar B,B_{i_2+1}).   
    \end{equation*}
    Spoiler chooses this set $B_{i_2+1}$. 
    Duplicator must answer with a set
    $A_{i_2+1}$ of size at least $l$. By the choice of $l$ we conclude that
    \begin{equation*}
      \Struc{A} \not\models \psi(\bar a, \bar A,A_{i_2+1}).      
    \end{equation*}
    By the inductive hypothesis, spoiler has a winning
    strategy in the resulting position for the $(k-1)$-round game. 
    
    For the other direction, assume that 
    $(\Struc{A},\bar a, \bar A)$ and $(\Struc{B}, \bar b, \bar B)$ 
    are indistinguishable by $\WMSOB$-formulas of quantifier rank $k$.
    Duplicator's strategy is as follows.
    \begin{itemize}
    \item If spoiler plays an element move choosing without loss of
      generality $a_{i_1+1}\in\Struc{A}$, let $\Phi$ be the  set of all
       $\WMSOB$-formulas $\varphi$ of quantifier rank $k-1$ such that 
      $\Struc{A} \models \varphi(\bar a, a_{i_1+1}, \bar A)$. Since $\Phi$ is finite up to logical
      equivalence, there is a $\WMSOB$-formula
      $\psi$ of quantifier rank $k-1$ such that $\psi\equiv \bigwedge_{\varphi\in\Phi} \varphi$. 
      By the assumption (indistinguishability up to quantifier rank $k$) and the fact that
      $\Struc{A} \models \exists x \psi(\bar a, x, \bar A)$ we conclude that 
      $\Struc{B} \models
      \exists x \psi(\bar b, x, \bar B)$. Hence, there is an element
      $b_{i_1+1}\in\Struc{B}$ such that
      $\Struc{B} \models
      \psi(\bar b, b_{i_1+1}, \bar B)$.
      Thus, duplicator can respond with $b_{i_1+1}$ and obtain a
      position for which he has a winning strategy by the induction
      hypothesis.
    \item If spoiler plays a set move, we use the same strategy as in
      the element move. We only have to replace the element $a_{i_1+1}$ by
      Spoiler's set $A_{i_1+1}$ and the first-order quantifier by a set
      quantifier.
    \item Assume that spoiler plays a bound move, choosing
      $\Struc{B}$ and bound $l\in\N$.  Let
      \begin{equation*}
        \Phi_A = \Set{ \varphi |  \text{rank}(\varphi) = k-1, 
        \forall {M\subseteq \Struc{A}} \left(\lvert M
        \rvert \geq l \Rightarrow \Struc{A}\not\models \varphi(\bar a, \bar
        A, M) \right) }.
      \end{equation*}
      Note that  $\Struc{A} \models \Bound X \varphi(\bar a, \bar A, X)$ for all
      $\varphi\in\Phi_A$.  Thus, 
      $\Struc{B} \models \Bound X \varphi(\bar b, \bar B, X)$ for all
      $\varphi\in \Phi_A$. Since
      $\Phi_A$ is finite up to equivalence 
      we can fix  a number $m\in \N$ that serves as a bound in
      $(\Struc{B},\bar b, \bar B)$ for all $\varphi\in \Phi_A$. Thus, for the set
      \begin{equation*}
        \Phi_B = \Set{ \varphi | \text{rank}(\varphi) = k-1,
        \forall {M\subseteq \Struc{B}} \left(\lvert M
        \rvert \geq m \Rightarrow \Struc{B}\not\models \varphi(\bar b,
        \bar B, M) \right) }
      \end{equation*}
      we have $\Phi_A\subseteq \Phi_B$.
      Duplicator answers Spoiler's challenge with this number $m$. 
      Then spoiler has to choose a set $B_{i_2+1}\subseteq \Struc{B}$
      of size at least $m$. 
      Let
      \begin{equation*}
        \Psi_B = \{ \varphi\mid \text{rank}(\varphi) = k-1, 
        \Struc{B}\models 
        \varphi(\bar b, \bar B, B_{i_2+1})\} .
      \end{equation*}
      Note that $\Phi_B \cap \Psi_B = \emptyset$. 
      Since $\Psi_B$ is finite up to equivalence, there is a
      $\WMSOB$-formula $\psi\in \Psi_B$ of quantifier rank $k-1$ such that 
      $\psi \equiv \bigwedge_{\varphi\in\Psi_B} \varphi$. In particular,
        $\psi\notin\Phi_B$.        
     Hence, $\psi \notin \Phi_A$ (since $\Phi_A \subseteq \Phi_B$). By the definition of 
     $\Phi_A$ this means that there is a  subset
        $A_{i_2+1}\subseteq\Struc{A}$ such that
        $\lvert A_{i_2+1}\rvert \geq l$ and $\Struc{A} \models
          \psi(\bar a, \bar A, A_{i_2+1})$. Duplicator chooses this set $A_{i_2+1}$.    
        The resulting position allows a winning strategy for
        duplicator by the induction hypothesis.   \qed
      \end{itemize}
 \end{proof}

\subsection{The Embeddable and the Unembeddable Triple-U-Structures}
\label{sec:TrippleUDefs}

In this section we define a class of finite structures $\Struc{G}_{n,m}$ for $n,m\in\N$.
Using these structures, we define for every $k \geq 0$ structures $\Struc{E}_k$ and $\Struc{U}_k$.
We show that for every $k \geq 0$, $\Struc{E}_k$ can be mapped homomorphically into a tree, whereas
$\Struc{U}_k$ cannot. In the next section, we will show that duplicator wins the $k$-round EF-game for 
both $\WMSOB$ and $\MSO$.

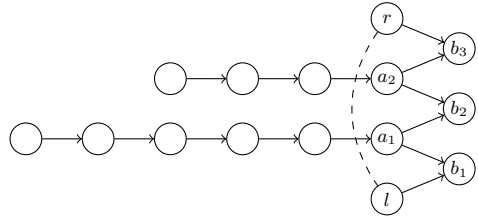
\begin{figure}[t]
  \centering
  
\begin{tikzpicture}
  [rotate = 90, transform shape, scale =0.8 ]   
  \begin{scope}
  [
  node distance = 1.2 and 0.5,
  on grid,
  basickonf/.style={ rotate= -90,
    circle, draw, solid, thin, minimum size = 15, inner sep = 0
  },
  konf/.style={ basickonf },
  less/.style= { <- },
  greater/.style= { -> },
  inc/.style= { dashed, bend right=35 },
  ]


  \foreach \pref in {0, ..., 0}
  \foreach \row in {0, ..., 0}
  {
    \node [konf] (\pref-\row-l) at ($(5*\pref, 4*\row)$) {$l$};
    \node [konf, below right = of \pref-\row-l] (\pref-\row-b_1)    {$b_1$}
    edge [less] (\pref-\row-l)
    ;
    \node [konf,  above right = of \pref-\row-b_1] (\pref-\row-a_0)    {$a_1$}
    edge [greater] (\pref-\row-b_1)
    ;
    \node [konf, below right  = of \pref-\row-a_0] (\pref-\row-b_2)    {$b_2$}
    edge [less] (\pref-\row-a_0)
    ;
    
    \node [konf, above right = of \pref-\row-b_2] (\pref-\row-c_0)    {$a_2$}
    edge [greater] (\pref-\row-b_2)
    ;
    \node [konf, below right  = of \pref-\row-c_0] (\pref-\row-b_3)    {$b_3$}
    edge [less] (\pref-\row-c_0)
    ;
    
    \node [konf, above right = of \pref-\row-b_3] (\pref-\row-r)    {$r$}
    edge [greater] (\pref-\row-b_3)
    edge [inc] (\pref-\row-l)
    ;

    \foreach \knoten / \lang in {a/5,c/3} {
      \foreach \i in {1, ..., \lang}
      {
        \pgfmathtruncatemacro{\minusOne}{\i-1};  
        \node [konf, above = of \pref-\row-\knoten_\minusOne] (\pref-\row-\knoten_\i) {}
        edge [greater] (\pref-\row-\knoten_\minusOne)
        ;
      }
    }
 }
  \end{scope}
 \end{tikzpicture}
  \caption{The standard $(5,3)$-triple-u, where we only draw the Hasse diagram for 
  $<^D$, and where dashed edges are $\inc$-edges.}
  \label{tab:35trippleU}
\end{figure}

  The \emph{standard plain triple-u} is the structure  $\Struc{P} =(P, <^P, \inc^P)$, where
  \begin{eqnarray*}
      P & =  & \{l,r,a_1,a_2,b_1,b_2,b_3\},\\
      <^P & = & \{ (l,b_1),(a_1,b_1),  (a_1,b_2), (a_2,b_2), (a_2,b_3),
      (r,b_3)\}, \text{ and}\\
      \inc^P & = & \{(l,r),(r,l)\} .
    \end{eqnarray*}
   We call a structure $(V,<,\inc)$ a \emph{plain triple-u} if it is
  isomorphic to the standard plain triple-u.
  For $n,m\in\N$,
  the \emph{standard $(n,m)$-triple-u}  is the
  structure $\Struc{G}_{n,m} = (D, <^D, \inc^D)$,
  where
  $$D = \{l,r,a_1,a_2,b_1,b_2,b_3\} \cup 
    (\{1, 2, \dots, n\}\times\{a_1\}) \cup
    (\{1, 2, \dots, m\}\times\{a_2\}), 
  $$
  and $<^D, \inc^D$ are the minimal relations such that
  \begin{itemize}
  \item   $\Struc{G}_{n,m}$ restricted to $\{l,r,a_1,a_2,b_1,b_2,b_3\}$ is
    the standard plain triple-u,
  \item 
    $(a_1, 1) < (a_1, 2) < \cdots < (a_1, n) < a_1$, and 
  \item
    $(a_2, 1) < (a_2, 2) < \cdots < (a_2, m) < a_2$. 
  \end{itemize}
  We call a graph $(V,<, \inc)$ an
  \emph{$(n,m)$-triple-u}  
  if it is isomorphic to the standard $(n,m)$-triple-u. 
  Figure \ref{tab:35trippleU} depicts a $(5,3)$-triple-u.

\begin{remark}
  For all $n,m\in\N$ and each $(n,m)$-triple-u  $\Struc{W}$ we fix an
  isomorphism $\psi_\Struc{W}$ between $\Struc{W}$ and the standard $(n,m)$-triple-u.
  Note that this isomorphism is unique if $n \neq m$. If $n = m$, then
  there is an automorphism of $\Struc{G}_{n,n}$ exchanging the nodes $l$ and $r$. Thus,
  choosing an isomorphism means to choose the left node of the triple-u. 
  For $x\in\{l,r,a_1,a_2,b_1,b_2,b_3\}$ we denote by $\Struc{W}.x$ the unique
  node $w \in \Struc{W}$ such that $\psi_\Struc{W}(w) = x$. 
  Furthermore, we call the linear order of size $n$ (resp., $m$) that consists of all proper
  ancestors of $\psi_\Struc{W}^{-1}(a_1)$ (resp., $\psi_{\Struc{W}}^{-1}(a_2)$) the \emph{left order} 
  (resp.,  \emph{right order}) of $\Struc{W}$. 
\end{remark}
Let $k\in\N$ be a natural number. Fix a strictly increasing
sequence $(n_{k,i})_{i\in\N}$
such that the linear order of length $n_{k,i}$ and the linear order of
length $n_{k,j}$ are 
equivalent with respect to $\WMSOB$-formulas of quantifier rank up
to $k$ for all $i,j\in\N$. 
Such a sequence exists because there are (up to equivalence) only
finitely many $\WMSOB$-formulas of quantifier rank $k$. 
Since the linear orders of length $n_{k,i}$ are finite,
they are equivalent with respect to both $\MSO$-formulas and
$\WMSO$-formulas of quantifier rank up to $k$.

\begin{definition}[The embeddable triple-u]
  Let $\Struc{E}_k$ be the  structure that consists of 
  \begin{enumerate}
  \item the disjoint  union of 
    infinitely many $(n_{k,1},n_{k,j})$-triple-u's and 
    infinitely many $(n_{k,j},n_{k,1})$-triple-u's   for all $j \geq 2$,
  \item one additional node $d$, and
  \item 
    for each triple-u $\Struc{W}$ an
    $<$-edge from $\Struc{W}.l$ to $d$.
  \end{enumerate}
  In the following we call $d$ the \emph{final node} of $\Struc{E}_k$
\end{definition}

\begin{lemma}
  For all $k\in\N$,  $\Struc{E}_k$ admits a homomorphism to a tree. 
\end{lemma}

\begin{proof}
  Using the procedure on the central points from the
  ordinal tree setting described in the proof of  Lemma~\ref{lem_ord_trees}, we first start
  adding the chains of each triple-u to the tree. In step $n_{k,1}$ we
  finally have placed all the chains of length $n_{k,1}$. Thus, for each
  triple-u $\Struc{W}$ either $\Struc{W}.a_1$ or $\Struc{W}.a_2$
  becomes central. Thus, in step 
  $n_{k,1}+1$ all the triple-u's split  into two disconnected
  components and the incomparability edges, which were avoiding that
  $\Struc{W}.l$ became central, now cease having such an effect.  
  We can therefore map $\Struc{W}.l$ at stage $n_{k,1}+2$ and the 
  final node $d$ in step $n_{k,1}+3$. Thus, it is easy to prove that the
  fixpoint procedure from the proof of Lemma~\ref{lem_ord_trees} terminates at
  stage $\omega$. Whenever this happens, the given structure admits a
  homomorphism to a tree, see Remark~\ref{rem-ordinal-tree-proof}.
  \qed
\end{proof}

\begin{definition}[The unembeddable triple-u]
  Let $\Struc{U}_k$ be the  structure that consists of 
  \begin{enumerate}
  \item the disjoint  union of 
    infinitely many $(n_{k,j},n_{k,j})$-triple-u's    for all $j \geq 2$,
  \item one additional node $d$, and
  \item 
    for each triple-u $\Struc{W}$ an
    $<$-edge from $W.l$ to $d$.
  \end{enumerate}
  In the following we call $d$ the \emph{final node} of $\Struc{U}_k$
\end{definition}

\begin{lemma}
  For all $k\in\N$,  $\Struc{U}_k$ does not admit a homomorphism to a tree. 
\end{lemma}

\begin{proof}
  Again, we consider the fixpoint procedure from the proof of Lemma~\ref{lem_ord_trees}.
  Assume that $\Struc{U}_k$ admits a homomorphism to a tree.
  Then, the final node $d$ has to be placed at some stage $i$ into the tree, i.e., in the notation of the 
  proof of Lemma~\ref{lem_ord_trees}, $d$ belongs to some set $C_i$ for $i < \omega$.
  But there is a $(n_{k,i},n_{k,i})$-triple-u $\Struc{W}$ and
  $\Struc{W}.l < d$.  Hence, $\Struc{W}.l$ has to be placed into the tree in one of the first $i-1$ stages.
  But $\Struc{W}.a_1$ and $\Struc{W}.a_2$ are the target nodes of chains of length $n_{k,i} \geq i$.
  Hence, after $i$ stages they are still not mapped into the tree. Therefore, after $i$ stages, $\Struc{W}.l$
  and $\Struc{W}.r$ are in the same connected component and they are linked by an $\inc$-edge.
  This contradicts the fact that $\Struc{W}.l$ was placed into the tree in one of the first $i-1$ stages.
  \qed
\end{proof}

\subsection{Duplicators Strategies in the $k$-Round Game}
\label{sec:Strategy}

We show that $\Theta$ does not have $\EHD$ by showing that duplicator
wins the $k$-round \MSO-EF-game and \WMSOB-EF-game on the pair
$(\Struc{E}_k, \Struc{U}_k)$ for each $k\in\N$. Hence, the two
structures are not distinguishable by $\Bool(\MSO, \WMSOB)$-formulas
of quantifier rank $k$. For $\MSO$ this is rather simple. Since the
linear orders of length $n_{k,i}$ and $n_{k,j}$ are indistinguishable
up to quantifier rank $k$, it is straightforward to compile the
strategies on these pairs of paths into a strategy on the whole
structures for the $k$-round game.  It is basically the same proof as
the one showing that a strategy on a pair $(\biguplus_{i \in I}
\Struc{A}_i, \biguplus_{i \in I} \Struc{B}_i)$ of disjoint unions can
be compiled from strategies on the pairs $(\Struc{A}_i, \Struc{B}_i)$.
In our situation there is an $i \in I$ such that $\Struc{A}_i =
\Struc{B}_i$ consists of infinitely many plain triple-u's together
with the final node, and the other pairs $(\Struc{A}_j, \Struc{B}_j)$
for $j \in I \setminus \{i\}$ consist of two linear orders that are
indistinguishable by $\MSO$-formulas of quantifier rank $k$.  We leave
the proof details as an exercise for the interested reader.

Compiling local strategies to a global strategy in the
$\WMSOB$-EF-game is much more difficult because strategies are not
closed under infinite disjoint unions. For instance, let $\Struc{A}$
be the disjoint union of infinitely many copies of the linear order of
size $n_{k,1}$ and $\Struc{B}$ be the disjoint union of all linear
orders of size $n_{k,j}$ for all $j\in\N$. Clearly, duplicator has a
winning strategy in the $k$-round game starting on the pair that
consists of the linear order of size $n_{k,1}$ and the linear order of
size $n_{k,j}$.  But in $\Struc{A}$ every linear suborder has size
bounded by $n_{k,1}$, while $\Struc{B}$ has linear suborders of
arbitrary finite size. This difference is of course expressible in
$\WMSOB$.
 Even though strategies in $\WMSOB$-games are not
closed under disjoint unions,  we can obtain a
composition result for disjoint unions on certain restricted
structures as follows. Let $\Struc{A} = \biguplus_{i\in\N} \Struc{A}_i$
and $\Struc{B} = \biguplus_{i\in\N} \Struc{B}_i$ be disjoint unions of
structures $\Struc{A}_i$ and $\Struc{B}_i$ satisfying the following
conditions:
\begin{enumerate}
\item All $\Struc{A}_i$ and $\Struc{B}_i$ are finite structures.
\item For every $i\in\N$, duplicator has a winning strategy in the
  $k$-round  $\MSO$-EF-game on $\Struc{A}_i$ and $\Struc{B}_i$.
\item \label{DisUnioncond_factormove} There is a constant
  $c\in\N\setminus\{0\}$
  such that whenever spoiler starts 
  the $\MSO$-EF-game on $(\Struc{A}_i, \Struc{B}_i$) with a set move
  choosing a set of size $n$ in 
  $\Struc{A}_i$ or $\Struc{B}_i$, then duplicator's strategy answers
  with a set of size at least $\frac{n}{c}$. 
\end{enumerate}
In this case duplicator has a winning strategy in the
$k$-round $\WMSOB$-EF-game on $\Struc{A}$ and $\Struc{B}$. 
To substantiate this claim, we sketch his strategy. For an element or
set move, duplicator 
just uses the local strategies from the $\MSO$-game to give an answer
to any challenge. For a bound move, duplicator does the following. If
spoiler's chooses the bound  $l\in\N$,  then duplicator chooses the
number $m$, which is the total number of elements in all substructures
$\Struc{A}_i$ or $\Struc{B}_i$ in which some elements have been chosen
in one of the previous rounds plus $c\cdot l$. This forces spoiler to
choose $c\cdot l$ elements in fresh substructures. Then duplicator
uses his strategy in each local pair of structures to give an answer
to spoiler's challenge. Since spoiler chose $c\cdot l$ elements in
fresh substructures, duplicator answers with at least 
$\frac{c\cdot l}{c}=l$ many elements in fresh substructures. 
This is a valid move and it preserves the existence of local winning
strategies between each pair $(\Struc{A}_i, \Struc{B}_i)$ for the
rounds yet to play.

From now on, we consider a fixed number $k\in\N$ and the game on the 
structures $\Struc{E}_k$ and $\Struc{U}_k$. We use a variant of the
closure under restricted disjoint unions, sketched above, to provide a winning strategy
for duplicator. In order to reduce
notational complexity we just write $\Struc{E}$ for $\Struc{E}_k$,
$\Struc{U}$ for $\Struc{U}_k$ and $n_i$ for $n_{k,i}$ (for all
$i\in\N$).  With $\bar E$ (resp. $\bar U$) we denote the set of all
maximal subgraphs that are $(n,m)$-triple-u's occurring in $\Struc{E}$
(resp., $\Struc{U}$) where $n$ and $m$ range over $\N$.  Note that
$\Struc{E}$ is the disjoint union of all $W \in \bar E$ together with
the final node, and similarly of $\Struc{U}$.
Unfortunately, we cannot apply the result on restricted disjoint
unions directly because of the
following problems.
\begin{itemize}
\item Due to the final nodes of $\Struc{E}$ and $\Struc{U}$, the
  structures are not disjoint unions of triple-u's. But since the
  additional structure in both structures is added in a uniform way
  this does not pose a problem for the proof.
\item The greater cause for trouble is that there is no constant $c$
  as in condition
  \ref{DisUnioncond_factormove} that applies uniformly to all  $\MSO$-EF-games
  on an $(n_j,n_1)$-triple-u of $\Struc{E}$ and an $(n_j,n_j)$-triple-u
  of $\Struc{U}$ for all
  $j\in\N$. The problem is that if spoiler chooses in his first move 
  all elements of the right order of the $(n_j,n_j)$-triple-u, then the only
  possible answer of duplicator is to choose the set of the $n_1$ many
  elements of the right order of the $(n_j, n_1)$-triple-u. 
  But since the numbers $n_j$ grow unboundedly, there is not constant $c$ such that the inequation
  $n_1\geq c n_j$ holds for all $j$.
  This problem does not exist for moves where spoiler chooses many
  elements in the left order of the $(n_j,n_j)$-triple-u. Duplicator's
  strategy allows to exactly choose the same subset of the left order
  of the $(n_j,n_1)$-triple-u. 
  This allows to overcome the problem that duplicator should answer
  challenges where spoiler chooses a large set with an equally
  large set (up to some constant factor): Instead of assigning
  each triple-u in $\bar E$ a fixed corresponding triple-u in $\bar U$, we  
  do this dynamically. If spoiler chooses a lot of elements from the
  left order of a fresh $(n_j, n_j)$-triple-u, then duplicator answers this
  challenge in a $(n_j,n_1)$-triple-u and we consider these two structures
  as forming one pair of the disjoint unions. On the other hand, if
  spoiler chooses a lot of elements from the right order of a fresh
  $(n_j,n_j)$-triple-u, then duplicators corresponding structure is chosen to be a
  fresh $(n_1,n_j)$-triple-u. In any case duplicator's local winning
  strategy may copy most of spoiler's choice (i.e., all elements
  chosen from the plain triple-u and from the order of length $n_j$
  from which spoiler has chosen more elements), thus producing a set
  which is at least half as big as spoiler's challenge.
\end{itemize}
In our prove we  encode this dynamic choice of corresponding
structures as a partial map $\varphi:\bar E \to \bar U$. The following
definition of a locally-$i$-winning position describes the requirements
on a position obtained after playing some rounds that allow to further use local
winning strategies in order to compile a winning strategy for the 
next $i$-rounds. It basically requires that the map $\varphi$ is such
that for each triple-u $W\in \dom(\varphi)$ the restriction of the
current position to $W$ and $\varphi(W)$ is a valid position in the
$i$-round $\WMSOB$-EF-game on $(W, \varphi(W))$ which is winning for
duplicator and that $\dom(\varphi)$ and $\image(\varphi)$ covers all 
elements that have been chosen so far (in an element move or as a
member of some set). 

\begin{definition}
  A position
  \begin{equation*}
    p = (\Struc{E}, e_1, \dots, e_{i_1}, E_1, \dots, E_{i_2},
    \Struc{U}, u_1, \dots, u_{i_1}, U_1, \dots, U_{i_2})    
  \end{equation*}
  in the $\WMSOB$-EF-game on $(\Struc{E}, \Struc{U})$ is called
  locally-$i$-winning (for duplicator) if there is a partial bijection
  $\varphi:\bar E \to \bar U$ such that 
  \begin{itemize}
  \item $\dom(\varphi)$ is finite,
  \item for all $W\in \bar E$, $W'\in \bar U$, and $1\leq j \leq i_1$,
    \begin{enumerate}
    \item if $e_j \in W$ then $W\in\dom(\varphi)$ and $u_j \in
      \varphi(W)$, and
    \item if $u_j \in W'$ then $W' \in \image(\varphi)$ and $e_j \in
      \varphi^{-1}(W')$,
    \end{enumerate}
  \item for all $W\in \bar E$, $W'\in \bar U$, and $1\leq j \leq i_2$,
    \begin{enumerate}
    \item if $E_j\cap W \neq \emptyset$ then $W\in\dom(\varphi)$ and
    \item  if $U_j\cap W' \neq \emptyset$ then $W'\in \image(\varphi)$, and
    \end{enumerate}
  \item 
    $\varphi$ is compatible with local strategies in the following sense:
    \begin{enumerate}
    \item For all $W\in\dom(\varphi)$,
      $x\in\{l,r,a_1,a_2,b_1,b_2,b_3\}$, $1\leq j \leq i_1$ and  $1\leq k \leq i_2$ we have
      \begin{itemize}
      \item $e_j = W.x  \Leftrightarrow u_j = \varphi(W).x$, and
      \item $W.x \in E_k \Leftrightarrow \varphi(W).x\in U_k$.
      \end{itemize}
    \item For all $W\in\dom(\varphi)$ and $1\leq j \leq i_1$, $e_j$ belongs to the left (resp., right)
    order of $W$ if and only if $u_j$ belongs to the  left (resp., right)
    order of $\varphi(W)$.
    \item For each $W\in\dom(\varphi)$, the restriction of the
      position $p$ to the left (resp., right) order
      of $W$ and the left (resp., right) order of
      $\varphi(W)$ is a winning position for duplicator in the
      $i$-round \WMSO-EF-game.
    \item For all $1\leq j \leq
      i_1$, $e_j$ is the final node of  $\Struc{E}$ if and only if
      $u_j$ is the final node of $\Struc{U}$.
    \item For all $1\leq j \leq
      i_2$, $E_j$ contains the final node of $\Struc{E}$ if and only if
      $U_j$ contains the final node of $\Struc{U}$.
  \end{enumerate}
  \end{itemize}
\end{definition}

\begin{remark}
  Note that the $\WMSOB$-EF-game on $(\Struc{E}, \Struc{U})$ starts in
  a locally-$k$-winning position where the partial map $\varphi$ is
  the map with empty domain. 
  Moreover, for all $i\in\N$, every locally-$i$-winning position
  is a winning position for duplicator in the $0$-round
  $\WMSOB$-EF-game.
\end{remark}

\begin{proposition}
  Duplicator has a winning strategy in the $k$-round $\WMSOB$-EF-game
  on $(\Struc{E}_k, \Struc{U}_k)$. 
\end{proposition}
Due to the previous remark, the proposition follows directly
form the following lemma.

\begin{lemma}
  Let $1\leq i \leq k$ be a natural number and
  $p$  a locally-$i$-winning position. Duplicator can respond
  any challenge of spoiler such that the next position is
  locally-$(i-1)$-winning.  
\end{lemma}

\begin{proof}
Let $\varphi: \bar E \to \bar U$ be the partial bijection for the locally-$i$-winning position $p$.
In the following, we say that an $(n,m)$-triple-u is \emph{fresh} if
it does not belong to $\dom(\varphi)\cup\image(\varphi)$. 
We consider the three possible types of moves for spoiler.
  \begin{enumerate}
  \item If spoiler plays an element move, there are the following possibilities.
    \begin{itemize}
    \item If spoiler chooses the final node of one of the structures,
      duplicator answers with the final node of the other.
    \item If spoiler chooses some node from an $(n,m)$-triple-u
      $W\in\dom(\varphi)$, then the local strategies allow duplicator
      to answer this move with a node from $\varphi(W)$.
    \item Analogously, if spoiler chooses some node from an $(n,m)$- triple-u
      $W\in\image(\varphi)$, then the local strategies allow
      duplicator to answer this move with a node from
      $\varphi^{-1}(W)$.
    \item If spoiler chooses a node from a fresh $(n,m)$-triple-u $W$ then
      duplicator can choose some fresh $(n',m')$-triple-u $W'$ from the other
      structure and can use the $\WMSO$-equivalence up to quantifier
      rank $k$ of the left and right orders of $W$ and $W'$ to
      find a response to spoilers challenge such that adding $(W,W')$
      (or $(W',W)$ depending on whether $W\in\bar E$) to $\varphi$
      leads to a locally-$(i-1)$-winning position. 
    \end{itemize}
  \item If spoiler plays a set move, then he chooses a finite set containing
    elements from some of the triple-u's from $\dom(\varphi)$ or
    $\image(\varphi)$ and from $l$ many fresh triple-u's. Choosing $l$
    fresh triple-u's from the other structure, we can find a response
    on each of the triple-u's corresponding to the local
    strategy similar to the case of the element move.
    The union of all these local responses is a response for
    duplicator that leads to a locally-$(i-1)$-winning position.
  \item If spoiler plays a bound move, we distinguish which structure
    he chooses.
    \begin{itemize}
    \item If he chooses structure $\Struc{U}$ and the bound $l\in\N$,
      let $Z_n$ be the (finite) set of all $(n,n)$-triple-u's
      occurring in $\image(\varphi)$ and set
      \begin{equation*}
        m_1 = \sum_{n\in\N}\sum_{W \in Z_n} (2n+7).
      \end{equation*}
      Duplicator responds with the bound $m = m_1 + 2l$.  Note that
      $2n+7$ is the size of an $(n,n)$-triple-u. Hence $m_1$ is the
      number of nodes in non-fresh triple-u's of $\Struc{U}$.  Assume
      that spoiler chooses some finite subset $S$ of $\Struc{U}$ with
      $\lvert S \rvert \geq m$. We construct a subset $S'$ in
      $\Struc{E}$ such that the resulting position is
      locally-$(i-1)$-winning. Moreover, we guarantee that for any
      fresh triple-u $W \in \bar U$ such that $S \cap W \neq
      \emptyset$, duplicator's response $S' \cap W'$ in a
      corresponding fresh triple-u $W' \in \bar E$ contains at least
      $\frac{1}{2}|S \cap W|$ many elements.  If $W_1, \ldots, W_z \in
      \bar U$ are all the fresh triple-u's that intersect $S$ non
      trivially, then we already argued that $|\bigcup_{i=1}^z (W_i
      \cap S)| \geq m-m_1 = 2l$.  Hence, duplicator's response $S'$
      contains at least $l$ many elements as desired.  The concrete
      choice of $S'$ is done as follows.
      \begin{enumerate}[(a)]
      \item For all $W \in\image(\varphi)$, duplicator chooses a set
        $S'_W\subseteq \varphi^{-1}(W)$ such that $S'_W$ is the answer
        to Spoiler's challenge $S \cap W$ according to a winning
        strategy in the $i$-round $\WMSO$-EF-game on the restriction
        of $p$ to $\varphi^{-1}(W)$ and $W$. This winning strategy
        exists because position $p$ is locally $i$-winning.
      \item Now consider a fresh $(n,n)$-triple-u $W \in \bar U$ with
        $W \cap S \neq \emptyset$.  Let $L$ (resp., $R$) be the nodes
        in the left (resp., right) order of $W$.  If $|L \cap S| \geq
        |R \cap S|$, then take a fresh $(n,n_1)$-triple-u $W' \in \bar
        E$ (note that $n \geq n_1$) and extend the partial bijection
        $\varphi$ by $\varphi(W') = W$. Duplicator chooses the subset
        $S'_W = \psi(S \cap W \setminus R)\cup T$, where $\psi$ is the
        obvious isomorphism between the $(n,0)$-sub-triple-u of $W$
        (i.e., $W \setminus R$) and the $(n,0)$-sub-triple-u of $W'$,
        and $T$ is an answer to spoilers move $S \cap R$ according to
        a winning strategy in the $i$-round $\WMSO$-EF-game between
        the right order of $W'$ and the right order of $W$. Note that
        $\lvert S'_W \rvert \geq \frac{1}{2} \lvert S \cap W \rvert$.
                
        If $|L \cap S| < |R \cap S|$, then let $W'$ be an $(n_1,n)$
        triple-u and use the same strategy but reverse the roles of
        the left and the right order of the chosen triple-u's.
      \item If the final node of $\Struc{U}$ is in $S$, let $S_d'$ be
        the singleton containing the final node of $\Struc{E}$,
        otherwise let $S_d'=\emptyset$.
      \end{enumerate}
      Finally, let $S'$ be the union of $S'_d$ and all sets $S'_W$
      defined in (a) and (b) above.  Since spoiler has chosen at least
      $2l-1$ many elements from fresh triple-u's, we directly conclude
      that $\lvert S' \rvert \geq l$. Moreover, since all the parts of
      $S'$ were defined using local strategies, we easily conclude
      that the position reached by choosing $S'$ is
      locally-$(i-1)$-winning.
    \item If spoiler chooses structure $\Struc{E}$ and bound $l\in\N$,
      we use a similar strategy. Let $Y_n$ be the set of all
      $(n_1,n)$-triple-u's and all $(n,n_1)$-triple-u's occurring in
      $\dom(\varphi)$, and define
      \begin{equation*}
        m_1 = \sum_{n\in\N} \sum_{W \in Y_n}  n_1+n+7,
      \end{equation*}
      and $m_2 = l\cdot n_1$. Note that $m_1$ is the number of nodes
      from non-fresh triple-u's from $\Struc{E}$.  Duplicator responds
      with $m = m_1 + m_2 + l$.  Let $S \subseteq \Struc{E}$ be
      Spoiler's set with $\lvert S\rvert\geq m$. There are at least
      $m_2+l$ elements in $S$ chosen from fresh triple-u's $W_1, W_2,
      \dots, W_z \in \bar E$. Either $z>l$ or spoiler has chosen at
      least $l$ elements from $W_1\cup W_2 \cup \dots\cup W_z$ that do
      not belong to the orders of length $n_1$ (which in total contain
      only $z \cdot n_1 \leq l \cdot n_1 = m_2$ many elements).
      Duplicator chooses his response $S'$ in $\Struc{U}$ as follows:
      \begin{enumerate}[(a)]
      \item For all $W \in\dom(\varphi)$, duplicator chooses a set
        $S'_W\subseteq \varphi(W)$ such that $S'_W$ is the answer to
        Spoiler's challenge $S \cap W$ according to a winning strategy
        in the $i$-round $\WMSO$-EF-game on the restriction of $p$ to
        $W$ and $\varphi(W)$. This winning strategy exists because
        position $p$ is locally $i$-winning.
      \item Now consider a fresh triple-u $W \in \bar E$ with $W \cap
        S \neq \emptyset$.  If $W$ is an $(n_1, n)$-triple-u or an
        $(n, n_1)$-triple-u, let $W'\in \bar U$ be a fresh $(n,
        n)$-triple-u of 
        $\Struc{U}$, and extend the partial bijection $\varphi$ by
        $\varphi(W) = W'$. Let us consider the case that $W$ is an
        $(n, n_1)$-triple-u (for the other case one can argue
        analogously) and let $R$ be the right order (of size $n_1$) of
        $W$.  Duplicator chooses the subset $S'_W = \psi(S \cap W
        \setminus R)\cup T$, where $\psi$ is the obvious isomorphism
        between the $(n,0)$-sub-triple-u of $W$ (i.e., $W \setminus
        R$) and the $(n,0)$-sub-triple-u of $W'$, and $T$ is an answer
        to Spoiler's move $S \cap R$ according to a winning strategy
        in the $i$-round $\WMSO$-EF-game between the right order of
        $W$ and the right order of $W'$. We can assume that $S'_W \neq
        \emptyset$. because we have $S \cap W \setminus R \neq
        \emptyset$ or $S \cap R \neq \emptyset$ and in the latter case
        $T$ can be chosen to be non-empty.
     \item If the final node of $\Struc{E}$ is in $S$, let $S_d'$ be
        the singleton containing the final node of $\Struc{U}$,
        otherwise let $S_d'=\emptyset$.
      \end{enumerate}
      Finally, let duplicator's response $S'$ be the union of $S'_d$
      and all sets $S'_W$ defined in (a) and (b) above.  By the
      argument before (a), duplicator selects in (b) in total at least
      $l$ elements.  Moreover, since all the parts of $S'$ where
      defined using local strategies, we easily conclude that the
      position reached by choosing $S'$ is locally-$(i-1)$-winning.
      \qed
    \end{itemize}
  \end{enumerate}
\end{proof}

\section{Open Problems}

The main open problem that remains is whether 
the problem $\SAT(\Gamma^=)$ is decidable for the class $\Gamma$ of all trees (or equivalently,
the single infinite binary tree). We have only proved that the $\EHD$-method cannot yield 
decidability. Currently, we are investigating automata theoretic approaches to this question.

\subsection*{Acknowledgment}
We thank Manfred Droste for fruitful discussions on universal
structures and semi-linear orders.

\appendix

\section{Universal Semi-linear Order}
\label{sec:universalOrder}

In the following we define a semi-linear order which is universal for
the class of all countable semi-linear orders. The fact that this
order is universal is known to the experts in the field of semi-linear
orders. Unfortunately, to our best knowledge there is no proof of this
fact in the literature. Hence we provide a proof for this result.

\begin{definition}
  Let $\Struc{U}=(U, <, \inc)$ be the countable semi-linear order with:
  \begin{itemize}
  \item $U = (\N\Q)^*$,
  \item $<$ is the strict order induced by
    $n_1p_1n_2p_2 \cdots n_kp_k \leq m_1q_1m_2q_2\cdots m_lq_l$  iff
    \begin{itemize}
    \item $k\leq l$, $n_i=m_i$ for all $1\leq i \leq k$, 
      $p_i=q_i$ for all $1\leq i \leq k-1$ and $p_k \leq q_k$, and
    \item ${\inc} = {\inc_<}$. 
    \end{itemize}
  \end{itemize}
  We call $\Struc{U}$ the universal countable semi-linear order. 
\end{definition}
Note that
  Droste~\cite{Droste85} has already studied this and similar orders. 

For $u = n_1p_1n_2p_2 \cdots n_kp_k \in U$ (with $k\geq 1$) and $q\in
\Q$, we define
\begin{equation} \label{def-addition}
u+q = n_1p_1n_2p_2\cdots n_{k-1}p_{k-1}n_k(p_k+q) .
\end{equation}
 We say that a countable semi-linear order $(A, <, \inc)$ is closed under
finite infima if 
for each finite set $S \subseteq A$ the linear order 
$\{a\in A\mid  a\leq s \text{ for all }s\in S\}$ has a maximal element, which is 
denoted by $\inf(S)$. 
Let $E=(a_i)_{i\in \N}$ be a repetition-free enumeration of $A$. We say $E$ is closed
under infima if for each initial subset $A_i=\{a_1, a_2, \dots, a_i\}$
and each $S\subseteq A_i$ we have $\inf(S) \in A_i$. 

\begin{lemma} \label{lemma-inf}
  Let $\Struc{A} = (A, < ,\inc)$ be a countable semi-linear order. 
  There is a countable semi-linear order $\Struc{B}$ that is closed
  under finite infima and an injective homomorphism from 
   $\Struc{A}$ to $\Struc{B}$.
\end{lemma}

\begin{proof}
  For a nonempty subset $S\subseteq A$ we set 
  $\prefixCl S = \{a\in A \mid \forall{s\in S} (a \leq s)\}$. 
  Let $\bar A$ be the set of finite nonempty subsets of $A$, which 
  is obviously
  countable. We define an equivalence on $\bar A$ by setting 
  $S\sim T$ iff $\prefixCl S = \prefixCl T$.  For all $S\in \bar A$, $[S]$ denotes its
  equivalence class. Let $B$ be the set of all equivalence classes. 
  We define an order $\sqsubset$ on $B$ by
  $[S] \sqsubset [T]$ if and only if 
  $\prefixCl S \subsetneq \prefixCl T$. 
  
  We claim that $\Struc{B} = (B, \sqsubset, \inc_\sqsubset)$ is a
  semi-linear order that is closed
  under finite infima and that  the map $\phi$ given by $\phi(a) \mapsto
  [\{a\}]$ is an injective homomorphism of $\Struc{A}$ in $\Struc{B}$. 
  \begin{itemize}
  \item  
    $\Struc{B}$ is obviously a partial order. Moreover, note that
    $\prefixCl S$ is a linear and  downwards closed suborder of $\Struc{A}$
    for every nonempty finite set $S \subseteq A$. In order to show
    that $\Struc{B}$ is semi-linear, assume that 
    $S_1 \sqsubset S$ and $S_2 \sqsubset S$. 
    Thus, all elements from $\prefixCl S_1$ and all elements from 
    $\prefixCl S_2$ are comparable. Since both sets are downwards
    closed, this directly implies that either 
    $[S_1] = [S_2]$, $[S_1] \sqsubset [S_2]$ or 
    $[S_2] \sqsubset [S_1]$.
  \item Let us show that  $\Struc{B}$ is closed under finite infima: 
    Let  $S, S_1, \ldots, S_n$ be finite subsets of $A$ and assume that $[S] \sqsubseteq [S_i]$ for all $1 \leq i \leq n$.
    Thus, $\prefixCl S \subseteq \prefixCl S_i$. Hence, $\prefixCl S \subseteq \bigcap_{i=1}^n \prefixCl S_i = \prefixCl \bigcap_{i=1}^n S_i$.
    Since $\prefixCl \bigcap_{i=1}^n S_i  \subseteq \prefixCl S_i$ for all $1 \leq i \leq n$, $[\bigcap_{i=1}^n S_i] = \inf(\{ [S_1], \ldots [S_n] \})$.
  \item For $a,b\in A$ with $a\neq b$ we have $b\notin
    \prefixCl\{a\}$ or $a\notin \prefixCl\{b\}$. Thus, $\phi$ is an
    injective map from $A$ to $B$. 
    Moreover, $a < b$ implies $\prefixCl\{a\} \subsetneq
    \prefixCl\{b\}$, i.e., $\phi(a) \sqsubset \phi(b)$. 
    Similarly, $a\inc b$ implies $a\notin \prefixCl\{b\}$ and
    $b\notin \prefixCl\{a\}$, i.e.,  $\phi(a) \inc_\sqsubset
    \phi(b)$. \qed
  \end{itemize}
\end{proof}

\begin{lemma} \label{lemma-enumerate-inf}
  Let $\Struc{A} =( A, < , \inc)$ be a countable semi-linear order that
  is closed under finite infima. There is a repetition-free enumeration of
  $\Struc{A}$, which is closed under infima. 
\end{lemma}

\begin{proof}
  Fix an arbitrary repetition-free enumeration $(a_i)_{i\in\N}$ of $A$. 
  Assume that we have constructed a sequence $b_1, b_2, \dots, b_i$
  such that $B_j = \{b_1, b_2, \dots, b_j\}$ is closed under infima
  for every $j\leq i$. Let $k\in \N$ be minimal with $a_k\notin B_i$. 
  Let $b'_1 < b'_2 < \dots b'_m < a_k$ be the list of all infima of
  the form $\inf(S\cup\{a_k\})$ for $S\subseteq B_i$ that are not
  contained in $B_i$. This list is indeed linearly ordered by $<$ since
  all elements in the list are bounded by $a_k$. 
  Now set $b_{i+l} = b'_l$ for all $1 \leq l \leq m$ and set 
  $b_{i+m+1} = a_k$.  The resulting sequence $b_1, \ldots, b_{i+m+1}$
  contains $a_k$ and $B_j = \{b_1, b_2, \dots, b_j\}$ is closed under infima
  for every $j\leq i+m+1$. This can be easily shown using the fact that 
  $\inf(X \cup \{ \inf(Y)\}) = \inf(X \cup Y)$ for all sets $X$ and $Y$.
  
  Repeating this construction leads to an enumeration $(b_i)_{i\in\N}$
  of $A$ with the desired property. \qed
\end{proof}

\begin{lemma}
  Let $\Struc{A}=(A, \sqsubset, \inc_\sqsubset)$ be a countable
  semi-linear order. There exists an injective homomorphism from
  $\Struc{A}$ into $\Struc{U}$. 
\end{lemma}

\begin{proof}
 Due to Lemma~\ref{lemma-inf} and \ref{lemma-enumerate-inf}, we may assume that
 $\Struc{A}$ is  closed under finite infima and that
 $(a_i)_{i\in\N}$ is a repetition-free enumeration of $A$ which is closed under
 finite infima. Set $A_i=\{a_1, \dots, a_i\}$ and
 $\phi_1: A_1 \to U$ with $\phi_1(a_1) = 00 \in \N\Q$. 

 Inductively, we construct injective  homomorphisms $\phi_i:A_i\to U$ such that
 \begin{enumerate}
 \item $\phi_{i+1}$ extends $\phi_i$, and
 \item for all 
   $u = n_1p_1n_2p_2\dots n_{k}p_{k} \in \image(\phi_i)$ and all
   $1\leq j \leq k$ we have $p_j\in \frac{1}{2^i}\Z$. 
 \end{enumerate}
 Assume that $\phi_i$ has already been constructed. 
 We distinguish two cases.
 \begin{enumerate}
 \item If there is some $a\in A_i$ with $a_{i+1} \sqsubset a$ let
    $u = \inf \{a \in A_i\mid a_{i+1} \sqsubset a\}$. Note that $a_{i+1} \sqsubseteq u$.
    Since the enumeration is closed under infima, we have $u \in A_i$ (and thus $a_{i+1} \sqsubset u$)
     and we can define
    $\phi_{i+1}(a_{i+1}) = \phi_i(u)+ (\frac{-1}{2^{i+1}})$, where we add according to \eqref{def-addition}. 
    Note that $\phi_{i+1}(a_{i+1}) < \phi_i(u) = \phi_{i+1}(u)$.
    In order to prove that this defines a homomorphism, we distinguish
    the following cases:
    \begin{enumerate}
    \item If $a_{i+1} \sqsubset a$ for some $a\in A_i$ then 
      $ u \sqsubseteq  a$. Hence, 
      $\phi_{i+1}(a_{i+1}) < \phi_{i+1}(u) = \phi_i(u) \leq \phi_i(a) =
      \phi_{i+1}(a)$ as desired.
    \item If $a_{i+1} \sqsupset a$ for some $a\in A_i$, then $a
      \sqsubset u$. Hence, $\phi_{i+1}(a) = \phi_i(a) <
      \phi_i(u)$. Since $\phi_i$ uses 
      only rationals from $\frac{1}{2^i}\Z$, we conclude that
      $\phi_{i+1}(a) 
        \leq \phi_i(u) + \frac{-1}{2^i} < \phi_{i+1}(a_{i+1})$ 
      as desired.
    \item If $a_{i+1} \inc_\sqsubset a$ for some $a\in A_i$, then
      $a \inc_\sqsubset u$.
      By induction,
      $\phi_{i+1}(a) = \phi_i(a) \inc_< \phi_i(u) = \phi_{i+1}(u)$. 
      Thus, the assumption $\phi_{i+1}(a) \leq \phi_{i+1}(a_{i+1})$
      leads by transitivity of $\leq$ to the contradiction 
      $\phi_{i+1}(a) \leq \phi_{i+1}(u)$. 
      Similarly, the assumption $\phi_{i+1}(a) > \phi_{i+1}(a_{i+1})$
       yields
      $\phi_{i+1}(a) \geq \phi_{i+1}(a_{i+1})+\frac{1}{2^{i+1}} =
      \phi_{i+1}(u)$. 
      We conclude that $\phi_{i+1}(a) \inc_< \phi_{i+1}(a_{i+1})$ as
      desired. 
    \end{enumerate}
  \item Otherwise, for all $j \leq i$ we know that  $\inf\{a_j,a_{i+1}\}$
    is strictly below $a_{i+1}$ and hence belongs to $A_i$ (since the 
    enumeration is closed under infima). In particular, the set 
    $\{ a\in A_{i}\mid a< a_{i+1}\}$ is not empty. By semi-linearity,
    $u=\max\{ a\in A_{i}\mid a< a_{i+1}\}$ is well-defined. Since
    $\image(\phi_i)$ is finite, there is some $n\in\N$ such that
    $\varphi_i(u) n 0$ is incomparable to all elements from the set
    $\varphi_i(A_i\setminus \{a\in A_i\mid a\leq u\})$. 
    Extending $\phi_i$ by setting $\phi_{i+1}(a_{i+1}) = \varphi_i(u)n0$ is easily
    shown to be a homomorphism. 
 \end{enumerate}
 Finally, the limit of $(\phi_i)_{i\in\N}$ clearly defines an injective
 homomorphism from $\Struc{A}$ into $\Struc{U}$. \qed 
\end{proof}

\end{document}